\documentclass[%
 aip,
 amsmath,amssymb,
reprint,        
onecolumn       
]{revtex4-2}

\usepackage{placeins}

\usepackage{amssymb}
\usepackage{amsmath,textcomp} 
\usepackage{bm}
\usepackage{amstext}
\usepackage{booktabs}
\usepackage{hhline}

\usepackage{graphicx}
\usepackage{dcolumn}
\usepackage{bm}

\usepackage[utf8]{inputenc}
\usepackage[T1]{fontenc}
\usepackage{mathptmx}
\usepackage{hyperref}
\usepackage{etoolbox}
\usepackage{array}
\newcolumntype{L}[1]{>{\raggedright\arraybackslash}p{#1}}

\usepackage[nohyperlinks,nolist]{acronym}
\usepackage{pstricks}
\usepackage{transparent}
\usepackage{comment}
\usepackage{wasysym}
\usepackage{pst-3dplot}
\usepackage{pst-tree}
\usepackage{pst-gr3d}
\usepackage{tikz}
\usetikzlibrary{arrows,calc,positioning,shadows,shapes,shapes.geometric}
\usepackage{pgfplots}
\usepackage{tabularx}
\usepackage{colortbl}
\usepackage{mathtools}
\usepackage{diagbox}
\usepackage{multirow}
\usepackage{eurosym}
\pgfplotsset{compat=1.18}
\usepackage{siunitx}
\DeclareSIUnit{\px}{px}

\usepackage{import}

\usepackage{caption}
\usepackage{subcaption}

\makeatletter
\def\@email#1#2{%
 \endgroup
 \patchcmd{\titleblock@produce}
  {\frontmatter@RRAPformat}
  {\frontmatter@RRAPformat{\produce@RRAP{*#1\href{mailto:#2}{#2}}}\frontmatter@RRAPformat}
  {}{}
}%
\makeatother

\begin{document}

\title[]{UrbanFlow-3K: A Dataset of 3,000 Lattice-Boltzmann Simulations of Random Building Layouts}

\author{Hojin Lee}
\affiliation{Data-Driven Fluid Engineering (DDFE) Laboratory, Inha University, Incheon, Republic of Korea}

\author{Andreas Lintermann}
\affiliation{Jülich Supercomputing Centre (JSC), Forschungszentrum Jülich, Jülich, Germany}

\author{Sangseung Lee*}
\affiliation{Data-Driven Fluid Engineering (DDFE) Laboratory, Inha University, Incheon, Republic of Korea}
\thanks{Corresponding author}
\email[Corresponding author. ]{sangseunglee@inha.ac.kr}

\author{Mario Rüttgers*}
\affiliation{Data-Driven Fluid Engineering (DDFE) Laboratory, Inha University, Incheon, Republic of Korea}
\thanks{Corresponding author}
\email[Corresponding author. ]{m.ruettgers@inha.ac.kr}

\thanks{Mario Rüttgers and Sangseung Lee are corresponding authors.}

\date{\today}
\begin{abstract}
The analysis of flow around buildings has gained significant research interest across various domains, including pedestrian safety, pollutant dispersion, natural ventilation, and building energy efficiency. 
While these domains frequently include high-resolution computational fluid dynamics (CFD) data, 
predicting urban flow fields with machine learning (ML) models has emerged as a promising approach to overcome the prohibitive costs of CFD simulations. 
However, the availability of open-source datasets for training such ML models remains scarce. 
In particular, publicly available two-dimensional datasets of urban flow fields are nearly non-existent, 
despite their potential value for early development and debugging stages of data-driven models, 
before scaling to computationally expensive three-dimensional datasets.
To bridge this gap, this study presents a comprehensive dataset consisting of $3{,}000$ two-dimensional urban flow simulations conducted using a lattice-Boltzmann method across three distinct \textsc{Reynolds} numbers. 
The dataset contains the time-averaged velocity fields.
A key feature of this dataset is its high geometric diversity: each layout incorporates between three and six buildings with randomized sizes, positions, and rotation angles ranging from \ang{0} to \ang{90}. 
This extensive variability enables the dataset to capture several critical flow characteristics, including wake formation, flow acceleration, shielding effects, and recirculation zones, across a wide range of orchestrated urban canopies.
The large sample size and consistent simulation setup make the dataset particularly suitable for developing and benchmarking ML architectures. 
In addition, the dataset can support transfer-learning strategies in which models trained on large two-dimensional datasets are adapted to smaller and more computationally expensive three-dimensional datasets.

\vspace{0.8em}
\noindent\textbf{Keywords:} Computational fluid dynamics, Urban canopy modeling,  Urban geometric diversity, Training data generation

\end{abstract}

\maketitle
\preprint{AIP/123-QED}









\begin{acronym}
\acro{ID}[ID]{identifier}
\acro{CNN}[CNN]{convolutional neural network}
\acro{CFD}[CFD]{computational fluid dynamics}
\acro{DDP}[DDP]{data distributed parallel}
\acro{RANS}[RANS]{Reynolds-averaged Navier Stokes}
\acro{UAV}[UAV]{unmanned aerial vehicle}
\acro{PMM}[PMM]{point-mass model}
\acro{MLP}[MLP]{multilayer perceptron}
\acro{ML}[ML]{machine learning}
\acro{RL}[RL]{reinforcement learning}
\acro{EC}[EC]{EdgeConv}
\acro{GC}[GC]{GraphConv}
\acro{ESPP-RL}[ESPP-RL]{energy-saving path planning reinforcement learning}
\acro{SVM}[SVM]{support vector machine}
\acro{LTD}[LTD]{limited thrust decomposition}
\acro{GCNN}[GCNN]{graph convolutional neural network}
\acro{3D-PTV}[3D-PTV]{3D particle-tracking velocimetry}
\acro{ANN}[ANN]{artificial neural network}
\acro{BGK}[BGK]{Bhatnagar–Gross–Krook}
\acro{CFD}[CFD]{computational fluid dynamics}
\acro{CAD}[CAD]{Computer Aided Design}
\acro{CPU}[CPU]{central processing unit}
\acro{CNPAS}[CNPAS]{congenital nasal pyriform aperture stenosis}
\acro{CT}[CT]{computed tomography}
\acro{DICOM}{digital imaging and communications in medicine}
\acro{DNS}[DNS]{Direct Numerical Simulation}
\acro{GPU}[GPU]{graphics processing unit}
\acro{GIS}[GIS]{Geographic Information System}
\acro{GNN}[GNN]{graph neural network}
\acro{GUI}[GUI]{graphical user interface}
\acro{GA}[GA]{Genetic Algorithm}
\acro{GPR}[GPR]{Gaussian Process Regression}
\acro{HPC}[HPC]{High-Performance Computing}
\acro{IGA}[IGA]{isogeometric analysis}
\acro{JURECA-DC}[JURECA-DC]{Jülich Research on Exascale Cluster Architectures}
\acro{JSC}[JSC]{Jülich Supercomputing Centre}
\acro{JLESC}[JLESC]{Joint Laboratory for Extreme Scale Computing}
\acro{LB}[LB]{lattice-Boltzmann}
\acro{LBM}[LBM]{lattice-Boltzmann Method}
\acro{LES}[LES]{large-eddy simulation}
\acro{LiDAR}[LiDAR]{Light Detection and Ranging}
\acro{m-AIA}[m-AIA]{multiphysics Aerodynamisches Institut Aachen}
\acro{MAPE}[MAPE]{mean absolute percentage error}
\acro{MaxAPE}[MaxAPE]{maximum absolute percentage error}
\acro{MARME}[MARME]{miniscrew-assisted rapid maxillary expansion}
\acro{ML}[ML]{machine learning}
\acro{MSE}[MSE]{mean squared error}
\acro{MRV}[MRV]{Magnetic Resonance Velocimetry}
\acro{NetCDF}[NetCDF]{Network Common Data Form}
\acro{NURBS}[NURBS]{non-uniform rational B-splines}
\acro{PPDF}[PPDF]{particle probability distribution function}
\acro{PIV}[PIV]{Particle Image Velocimetry}
\acro{PV}[PV]{Photovoltaic}
\acro{PINN}[PINN]{Physics-Informed Neural Network}
\acro{PT}[PT]{Passage Time}
\acro{PA-CNN}[PA-CNN]{Physics-Aware Convolutional Neural Network}
\acro{PA-GNN}[PA-GNN]{Physics-Aware Graph Neural Network}
\acro{RANS}[RANS]{Reynolds-averaged Navier Stokes}
\acro{ROI}[ROI]{region of interest}
\acro{STL}[STL]{Stereolithography}
\acro{SP}[SP]{start point}
\acro{WP}[WP]{way point}
\acro{DE}[DE]{destination}
\acro{StB}[StB]{Shake-the-Box}
\acro{SGMS-GNN}[SGMS-GNN]{Subgraph Multi-Scale Graph Neural Network}
\acro{RL}[RL]{reinforcement learning}
\acro{ZFS}[ZFS]{Zonal Flow Solver}
\acro{UAM}[UAM]{Urban Air Mobility}
\acro{CLB}[CLB]{Cumulant Lattice-Boltzmann}
\acro{VAWT}[VAWT]{Vertical Axis Wind Turbine}
\acro{HAWT}[HAWT]{Horizontal Axis Wind Turbine}
\acro{ECMWF}[ECMWF]{European Centre for Medium-Range Weather Forecasts}
\acro{ZFS}[ZFS]{Zonal Flow Solver}
\acro{CLB}[CLB]{Cumulant Lattice-Boltzmann}
\acro{MRT}[MRT]{Multiple-Relaxation-Time}
\acro{GRIB}[GRIB]{GRIdded Binary}
\acro{PID}[PID]{proportional-integral-derivative}
\acro{LQR}[LQR]{linear-quadratic regulator}
\acro{NMPC}[NMPC]{nonlinear model predictive control}
\acro{PSO}[PSO]{Particle Swarm Optimization}
\acro{MGMPSO}[MGMPSO]{Modified Group Mutation PSO}
\acro{ELBS}{Extended Land Beaufort Scale}
\acro{PIV}{Particle Image Velocimetry}
\acro{NDI}{Neighbor Distance Imputation}
\acro{MICE}{Multiple Imputations by Chained Equations}
\acro{GAIN}{Generative Adversarial Imputation Network}
\acro{AME}{Average Mean Error}
\acro{ARV}{Average R-squared Value}
\acro{POD}{proper orthogonal decomposition}
\acro{FNO}{fourier neural operator}
\acro{PDE}{partial differential equation}
\acro{RDN}{deeper residual dense network}
\acro{GAN}{generative adversarial network}
\acro{ELBS}{Extended Land Beaufort Scale}
\acro{HDF5}[HDF5]{hierarchical data format version 5}
\end{acronym}

\section{Introduction}
\label{sec1}

With rapid urbanization presenting numerous challenges, 
\ac{CFD} has become an essential tool for investigating flow fields in urban environments.
One class of studies focuses on understanding how the flow interacts with the built environment.
Such investigations aim to assess local flow characteristics in the vicinity of buildings and streets, with applications including pedestrian level wind, pollutant dispersion, wind loads, building energy efficiency, and natural ventilation strategies.

For example, Pancholy et al.~\cite{Pancholy2021} investigated characteristics of flow around two parallel buildings, 
an upstream and a downstream one, 
separated by a distance $S$. 
The results showed that zones with high velocities that cause discomfort and unpleasantness were reduced in uniform canyons where $S$ equals the building height or in non-uniform canyons where the downstream building was shorter.
In another example, Tominaga et al.~\cite{Tominag2011} investigated pollutant dispersion in a three-dimensional street canyon using both \ac{RANS} simulations and \acp{LES}.
\ac{RANS} simulations model all turbulent fluctuations and provide only time-averaged flow quantities, whereas \ac{LES} approaches explicitly resolve the large, energy-containing turbulent structures while modeling only the smaller scales, enabling a more faithful representation of flow unsteadiness at increased computational cost.
The authors highlighted that an \ac{LES} provides crucial information on instantaneous concentration fluctuations, 
which cannot be captured by a \ac{RANS} simulation. 
Consequently, the results indicate that plume dispersion in street canyons is highly unsteady. 
Jon et al.~\cite{Jon2023} analyzed the impacts of wind direction on the ventilation of canyons with balconies. 
For this purpose, five wind directions ($\alpha$=\ang{0}, \ang{30}, \ang{45}, \ang{60}, \ang{90}) and three balcony locations (leeward-side, windward-side, and both-sides) were investigated. 
They recommended that the leeward-side balconies should be used in designing buildings whenever possible.

A second class of applications treats the urban flow field as a key external factor influencing the performance and safety of other systems operating within the urban canopy.
Here, the primary interest lies not in the modification of the built environment, but in how the existing flow conditions affect mobile or adaptive agents.
For example, Gu et al.~\cite{GU2026} utilized \ac{CFD} simulations to calculate urban wind fields and proposed a wind-aware path planning algorithm for \acp{UAV}.
They integrated wind information from the simulation results into the cost function of a conventional path planning algorithm. 
This wind-aware approach resulted in a $6.23\%$ increase in ground speed and reduced energy consumption by $7.69\%$ compared to an existing method. 
Similarly, Rienecker et al. generated wind fields for a typical European urban environment model using \acp{LES}, 
and optimized flight paths with a tailored path planning algorithm~\cite{Rienecker2023}. 
Comparing energy-optimal paths against shortest paths across $12$ different scenarios, 
they demonstrated energy-per-distance savings ranging from a minimum of $5.0\%$ to a maximum of $47.0\%$. 
These studies underscore the growing importance of understanding fluid flow around buildings and directly linking it to other systems such as an improved efficiency when operating \acp{UAV}.

While high-fidelity \ac{CFD} simulations enable detailed assessments of how urban flow fields influence the built environment or mobile systems such as \acp{UAV}, their computational cost often limits their direct use in real-time decision-making or large parametric studies.
To address these challenges, \ac{ML} models are increasingly trained on \ac{CFD}-generated datasets to act as fast surrogate models for urban flow field predictions.
Such surrogates enable a rapid evaluation of wind fields across different urban layouts and boundary scenarios.
Peng et al.~\cite{Peng2024} proposed a \ac{FNO} surrogate model, 
which is a deep learning model specialized in solving \acp{PDE}. 
The training data was generated using CityFFD~\cite{Mortezazadeh2022},
a specialized CFD model designed for fast simulations of urban microclimates,
for a domain in Niigata, Japan, characterized by an $800\si{m}$ square area and an elevation reaching up to $300\si{m}$. 
The \ac{FNO} model demonstrated high accuracy when tested against unseen conditions, 
yielding a maximum error of $5\%$. 
Zhang et al.~\cite{Zhang2022} employed a \ac{GAN}-based framework combining a generator and discriminator to predict instantaneous velocity fields around a single building from sparse surface pressure measurements. They created a blend of \ac{GAN} and \ac{MSE} losses, which demonstrated that the blended approach reproduces both large-scale flow structures and turbulent statistics more faithfully.

Building and validating robust \ac{ML} surrogates for urban flow fields requires large, diverse, and publicly accessible high-fidelity flow datasets that span a wide range of urban configurations and flow conditions.
In this context, Nazarian et al.~\cite{Nazarian2025} developed the public UrbanTALES dataset comprising $538$ urban layouts generated through \acp{LES}. 
The dataset is categorized into two primary groups: $224$ idealized configurations and $314$ realistic configurations. 
The idealized configurations incorporate both aligned and staggered building arrays, considering $8$ distinct packing densities and two wind directions. 
These layouts are further characterized by four types of building height arrangements: (1) uniform arrays with constant building height; (2) continuous configurations representing gradual height distributions; (3) clustered configurations featuring distinct groups of low and tall buildings; and (4) high-rise configurations characterized by extreme height contrasts between high-rise and low-rise blocks. 
The realistic configurations were selected from major global cities to capture diverse urban densities and significant vertical heterogeneities.
Another example of a publicly available dataset for urban flow fields is the CityTransformer dataset, which comprises three-dimensional \ac{LES} data of plume dispersion and wind fields over a realistic urban area, enabling surrogate model training and validation across large domains with time-varying flow statistics~\cite{Asahi2022}.

While these public datasets provide highly resolved flow fields in realistic three-dimensional environments,
they remain limited for early development and debugging stages of data-driven models.
Simplified two-dimensional datasets provide an alternative by enabling the generation of substantially larger training sets at a fraction of the computational cost, 
thereby supporting systematic development and evaluation of data-driven flow prediction methods before extending them to full three-dimensional urban environments.
Similar strategies are widely used in other machine learning domains, 
where models are first trained on large simplified datasets before being adapted to more complex tasks. 
For instance, deep neural networks in computer vision are commonly pre-trained on large generic image datasets and subsequently fine-tuned for more specialized applications~\cite{Kolesnikov2020}. 
Likewise, natural language processing models are typically pre-trained on large text corpora before being adapted to downstream tasks such as question answering or sentiment analysis~\cite{Peters2018}. 
Such strategies highlight the potential of reduced-complexity datasets as a stepping stone toward more computationally demanding three-dimensional flow predictions. 
However, publicly available datasets specifically designed for two-dimensional urban flow configurations are largely absent.

To address this gap, the present work introduces a complementary dataset consisting of $3{,}000$ two-dimensional urban neighborhood layouts simulated using a \ac{LB} method at three distinct \textsc{Reynolds} numbers.
Here, the \textsc{Reynolds} number is defined based on the mean inflow velocity, a characteristic geometric length scale, and the kinematic viscosity of air. It is, therefore, directly related to the mean flow rate.
The dataset emphasizes systematic geometric diversity by allowing the number of buildings to vary between three and six, with randomized building sizes, positions, and rotation angles ranging from \ang{0} to \ang{90}, thereby covering a broad range of effective wind directions and obstacle interactions. 
The large sample size, controlled variability, and uniform simulation setup make it particularly well suited for benchmarking, sensitivity analyses, and optimization studies.
In particular, the dataset enables systematic investigations of ML architectures, training strategies, and data requirements under controlled conditions, which are often difficult to perform with computationally expensive three-dimensional datasets. 
Furthermore, the dataset can support transfer-learning strategies in which models are pre-trained on large two-dimensional datasets before being adapted to smaller and more expensive three-dimensional urban flow datasets. 
In this sense, the dataset complements existing three-dimensional datasets by providing both a scalable testbed for early stages of algorithm development and a foundation for transferring learned representations to higher-fidelity three-dimensional simulations.

The manuscript is structured as follows. Section~\ref{sec2} describes the computational methods that have been used to generate the dataset. 
Section~\ref{sec3} provides a grid refinement study, validation, and analyses of example flow fields from the dataset for various \textsc{Reynolds} numbers and building configurations. 
Furthermore, the dataset's structure is described in detail to ease data handling.
Finally, in Section~\ref{sec4}, the main novelties and limitations of this dataset are summarized and discussed.

\section{Methodology}
\label{sec2}

In this section, the numerical methods for grid generation and conducting the numerical flow simulations are explained in Sec.~\ref{subsec1}.
Next, the computational domain of the dataset's samples is presented in Sec.~\ref{subsec2}.
Finally, Section~\ref{subsec3} presents the methodology for processing the dataset to facilitate its use in \ac{ML} model training.

\subsection{Numerical Methods}
\label{subsec1}

Unstructured hierarchical Cartesian grids are generated for each sample of the dataset using the massively parallel grid-generation capabilities of the \ac{m-AIA} simulation framework~\cite{Lintermann2014,m-AIA,Lintermann2020a}.
The grid topology follows an octree-based structure, constructed by recursively subdividing an initial bounding square that encloses the computational domain.
For two-dimensional configurations, each parent cell is split into four child cells, yielding a hierarchical grid with explicit parent–child relationships.
Cells that fall outside the physical flow domain, such as those located within solid obstacles, are removed to reduce computational cost and improve grid efficiency.

Cells on very coarse levels up to a user-defined minimum refinement level are removed from the tree leading to a forest of octrees. 
The root nodes of these trees are ordered with a Hilbert space-filling curve~\cite{Sagan1994} for domain decomposition. 
Cells on finer refinement levels are ordered using a Z-curve (Morton order)~\cite{Morton1966}.
Local refinement can be applied near boundaries or in designated regions of interest.
During parallel execution, the Hilbert ordering preserves spatial locality by assigning neighboring cells to the same or nearby processing units.
Combined with Morton ordering within subdomains, this approach enhances cache efficiency and minimizes inter-process communication.
The final grid can be written in parallel \ac{NetCDF} or \ac{HDF5} format using concurrent I/O routines, enabling scalable and efficient data output for large-scale simulations~\cite{Li2003,hdf5}.

Two-dimensional flow simulations are performed using the \ac{LB} solver implemented in \ac{m-AIA}.
The solver has been widely used for various flow applications, 
with recent studies including respiratory flow simulations~\cite{Ruettgers2024,Ruettgers2025b,Ruettgers2025} or the numerical analysis of cerebral aneurysms~\cite{Ruettgers2025a}.
The LB method solves a discretized form of the Boltzmann equation employing the Bhatnagar–Gross–Krook (BGK) collision operator~\cite{heTheoryLatticeBoltzmann1997}, expressed as
\begin{equation}
f_i(\boldsymbol{x} + \boldsymbol{\xi}_i \delta t, t + \delta t) - f_i(\boldsymbol{x}, t)
= -\omega \left( f_i(\boldsymbol{x}, t) - f^{eq}_i(\boldsymbol{x}, t) \right),
\end{equation}
where $\omega$ represents the collision frequency
and $f_i$ denotes the particle probability distribution functions evaluated at lattice sites $\boldsymbol{x}$ and propagated along discrete velocity directions with the molecular velocity vector $\boldsymbol{\xi}_i$ over a time $t$ and time step $\delta t$.

The discretization employs the standard $D2Q9$ scheme~\cite{qianLatticeBGKModels1992}, with $i \in {1,\dots,Q}$.
The discrete equilibrium distribution function is given by
\begin{equation}
f^{eq}_i
= w_i \rho \left(
1 + \frac{\boldsymbol{\xi}_i \cdot \boldsymbol{u}}{c_s^2}
+ \frac{1}{2}\left( \frac{\boldsymbol{\xi}_i \cdot \boldsymbol{u}}{c_s^2} \right)^2
- \frac{\boldsymbol{u} \cdot \boldsymbol{u}}{2 c_s^2}
\right),
\end{equation}
where $\rho$ is the fluid density, $\boldsymbol{u}$ the macroscopic velocity, $w_i$ are the lattice weights, and $c_s = 1/\sqrt{3}$ the isothermal speed of sound.

Macroscopic flow quantities are computed from the distribution functions via
\begin{align}
\rho &= \sum_{i=1}^{Q} f_i, \\
\rho \boldsymbol{u} &= \sum_{i=1}^{Q} \boldsymbol{\xi}_i f_i.
\end{align}
The static pressure is related to the density through the equation of state $p_s = c_s^2 \rho$.



\subsection{Computational Domain}
\label{subsec2}
\begin{figure}[]
\centering
\include{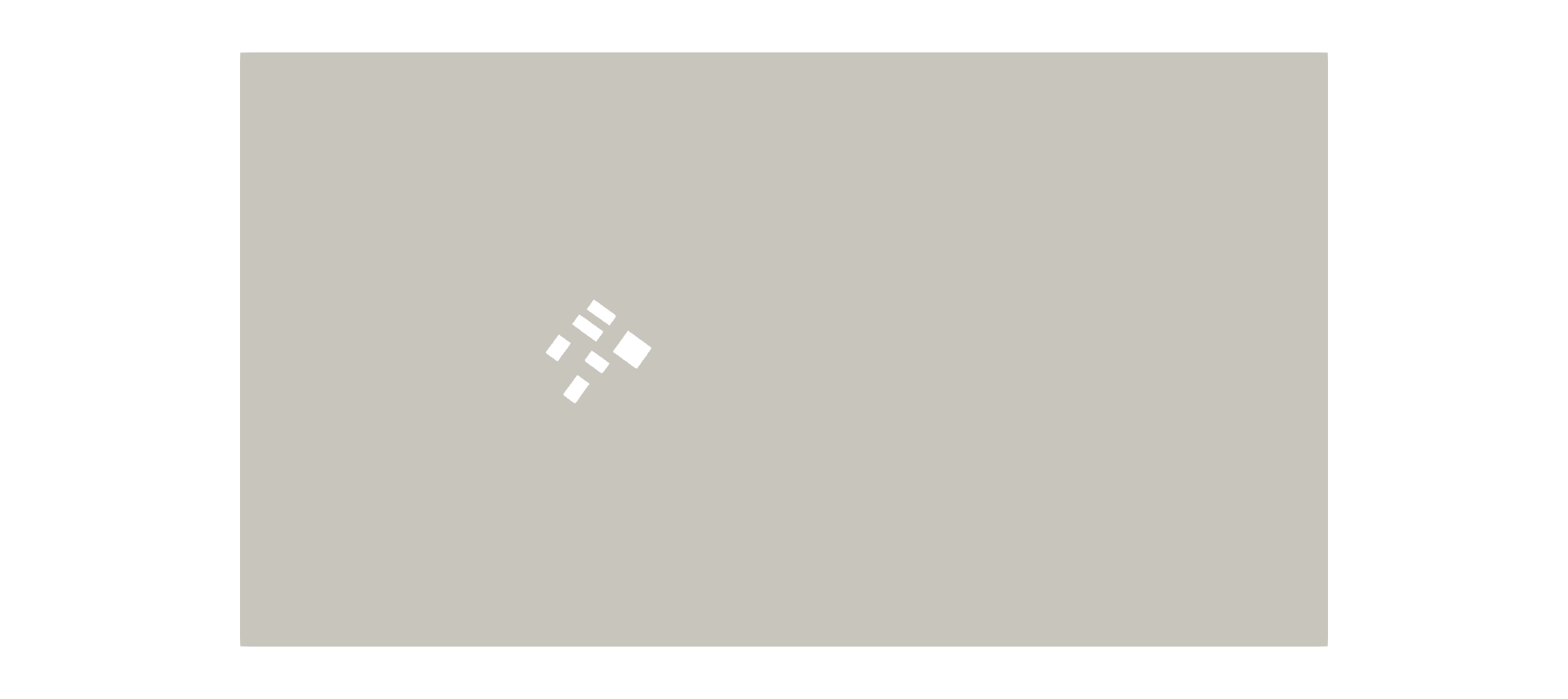}
\caption{Example for the simulation domain of a randomly generated urban layout.}
\label{fig:domain_ex}
\end{figure}

The dataset contains $3{,}000$ \ac{LB} simulations of flow through synthetically generated urban configurations.
An example configuration is illustrated in Fig.~\ref{fig:domain_ex}.
The characteristic length scale is defined as the minimum edge length of a building, denoted by $d_{ref} = 1$.
The computational domain spans a length of $L = 45d_{ref}$ and a height of $H = 30d_{ref}$.

Each urban configuration is generated procedurally using the following randomized parameters:
\begin{itemize}
\item A randomly selected number of non-overlapping buildings $B \in \{3,\ldots,6\}$, each with edge lengths $d$ sampled uniformly from the interval $d_{\text{ref}} < d < 2d_{ref}$.
\item Building centers $(x_{\text{cen}}, y_{\text{cen}})$ are placed randomly within a predefined \ac{ROI} bounded by $-2d_{ref} < x_{cen} < 4d_{ref}$ and $-3d_{ref} < y_{cen} < 3d_{ref}$.
\item The \ac{ROI} covers the area within $-3d_{ref} \leq x \leq 6d_{ref}$ and $-3.5d_{ref} \leq y \leq 3.5d_{ref}$.
\item The entire building ensemble is rotated about the center of the \ac{ROI} by a randomly sampled angle $0^\circ < \alpha < 90^\circ$.
\end{itemize}

The computational domain employs five boundary conditions, labeled BC~I through BC~V in Fig.~\ref{fig:domain_ex}.
At the inlet (BC~I), a uniform reference velocity $U_{ref}$ is prescribed, while the density is extrapolated from the interior cells.
At the outlet (BC~II), a constant reference density $\rho_{ref}$, and thus a fixed static pressure $p_{s,ref}$, is imposed, 
with the velocity extrapolated from the interior cells.
Slip-flow boundary conditions are applied at the top (BC~III) and bottom (BC~IV) boundaries.
At solid surfaces corresponding to building walls (BC~V), an interpolated bounce-back scheme is used to enforce the no-slip condition~\cite{bouzidiMomentumTransferBoltzmannlattice2001}.
In contrast to the no-slip condition, where particle distribution functions are reflected into the opposite lattice directions to impose zero wall velocity, the slip condition employs specular reflection.
Here, incoming distribution functions are assigned from the mirrored outgoing directions, eliminating the velocity component normal to the boundary while preserving the tangential component.

Each urban configuration is simulated at a prescribed \textsc{Reynolds} number defined as
\begin{equation}
Re = \frac{U_{ref} d_{ref}}{\nu},
\end{equation}
with $\nu$ denoting the kinematic viscosity.
The simulations cover representative flow regimes with $Re = 3{,}000$, $4{,}000$, and $5{,}000$, yielding a dataset with $1{,}000$ simulations per \textsc{\textsc{Reynolds}} number.

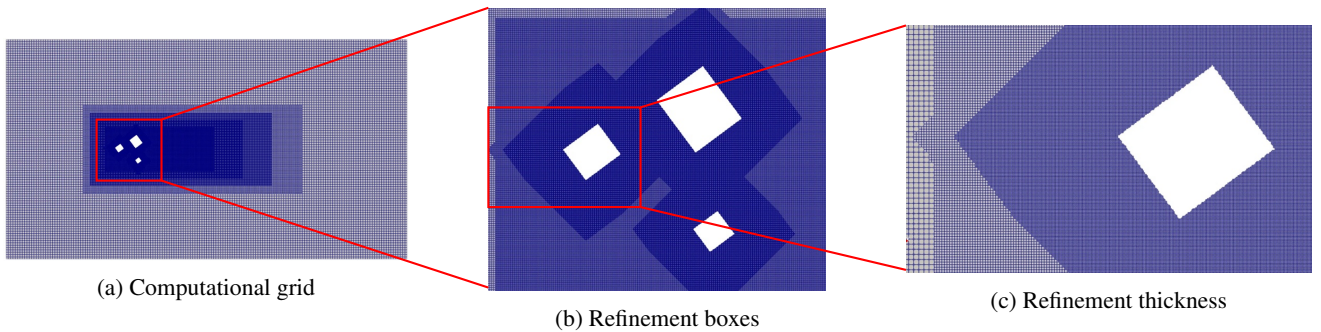
\begin{figure}[h]
    \centering
    \begin{tikzpicture}[remember picture]
        
        \node[inner sep=0] (main_node) at (0,0) {
            \begin{subfigure}[b]{0.3\textwidth}
                \centering
                \includegraphics[width=\textwidth]{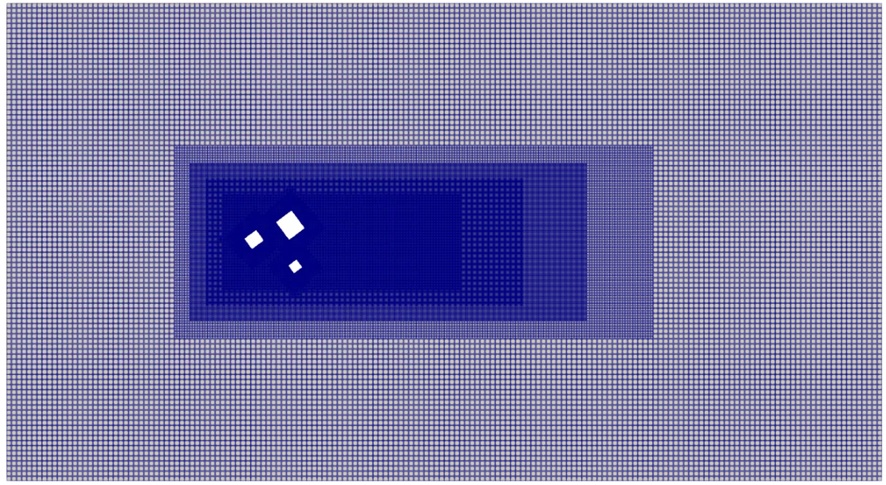}
                \caption{Computational grid}
                \label{fig:mesh_overall}
            \end{subfigure}
        };

        \node[inner sep=0] (zoom_node) at (6,0) {
            \begin{subfigure}[b]{0.25\textwidth}
                \centering
                \includegraphics[width=\textwidth]{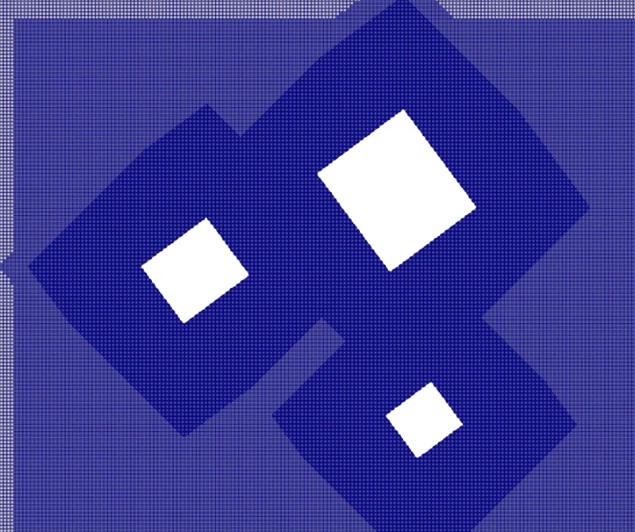}
                \caption{Refinement boxes}
                \label{fig:Intermediate zoom}
            \end{subfigure}
        };

        \node[inner sep=0] (closeup_node) at (12,0) {
            \begin{subfigure}[b]{0.3\textwidth}
                \centering
                \includegraphics[width=\textwidth]{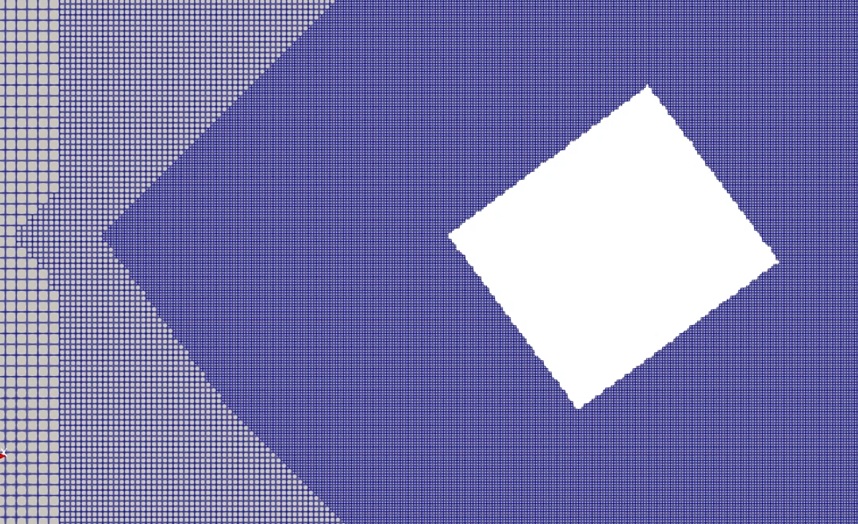}
                \caption{Refinement thickness}
                \label{fig:Close-up view}
            \end{subfigure}
        };

        \begin{scope}[shift={(main_node.south west)}, 
                      x={(main_node.south east)}, 
                      y={(main_node.north west)}, 
                      red, thick]
            \draw (0.23, 0.475) rectangle (0.39, 0.7) coordinate (roi1_top) at (0.39, 0.7) 
                                                     coordinate (roi1_bottom) at (0.39, 0.475);
            \draw (roi1_top) -- (zoom_node.north west);
            \draw (roi1_bottom) -- ([yshift=2.2em]zoom_node.south west);
        \end{scope}

        \begin{scope}[shift={(zoom_node.south west)}, 
                      x={(zoom_node.south east)}, 
                      y={(zoom_node.north west)}, 
                      red, thick]
            \draw (0, 0.4) rectangle (0.45, 0.7) coordinate (roi2_top) at (0.45, 0.7) 
                                                     coordinate (roi2_bottom) at (0.45, 0.4);
            \draw (roi2_top) -- (closeup_node.north west);
            \draw (roi2_bottom) -- ([yshift=2.2em]closeup_node.south west);
        \end{scope}

    \end{tikzpicture}
    
    \caption{Example for a grid around multiple buildings.}
    \label{fig:mesh_refinement_levels}
\end{figure}

Figure~\ref{fig:mesh_refinement_levels} visualizes the computational grid for a representative case with three buildings. 
In Fig.~\ref{fig:mesh_overall}, the arrangement of the rectangular refinement boxes used to locally increase the grid resolution around the urban geometries is illustrated. 
These refinement regions are 
extended vertically to adequately capture near-building flow features and wake development.
In Figs.~\ref{fig:Intermediate zoom} and~\ref{fig:Close-up view}, an additional boundary-layer–oriented refinement applied directly to the building surfaces is shown, 
ensuring sufficient resolution of near-wall gradients and shear layers.
The figures also show that the rectangular refinement boxes are locally extended if a building's local refinement is too close to them to guarantee a sufficiently large thickness at each refinement level.

\subsection{Generation of Training Data for \ac{ML} Models}
\label{subsec3}
The \ac{LB} simulations are conducted on the \ac{GPU} partition of the \ac{JURECA-DC} supercomputer at Forschungszentrum Jülich~\cite{JURECA,Krause2018}. Each node of the \ac{GPU} partition is equipped with four NVIDIA A100 \acp{GPU} and two AMD EPYC $7742$ \acp{CPU}, each featuring $64$ cores clocked at $2.25$ GHz, and $512$ GB of DDR4 memory.

Beyond capturing complex flow patterns and mitigating the lack of extensive urban flow datasets, these 3,000 simulations serve as a foundation for training diverse \ac{ML} architectures. 
For each simulation, all macroscopic flow variables are written as column vectors, 
where each entry corresponds to a single cell of the grid. 
The ordering of cells follows the previously explained combination of Hilbert space-filling and Morton (Z-order) curves.
While this representation is well suited for large-scale \ac{CFD} simulations and post-processing, it is not directly compatible with many \ac{ML} models.
In particular, a large class of \ac{ML} architectures, 
such as \acp{CNN}, 
assumes data defined on uniformly spaced, structured Cartesian grids. 
Other approaches, including \acp{GNN}, 
require explicit neighborhood information in the form of adjacency lists or matrices.
Furthermore, the presence of locally refined cells with varying spatial resolution complicates the direct use of the raw data in many \ac{ML} pipelines, 
as the majority of learning algorithms implicitly assumes a single, uniform spatial scale.

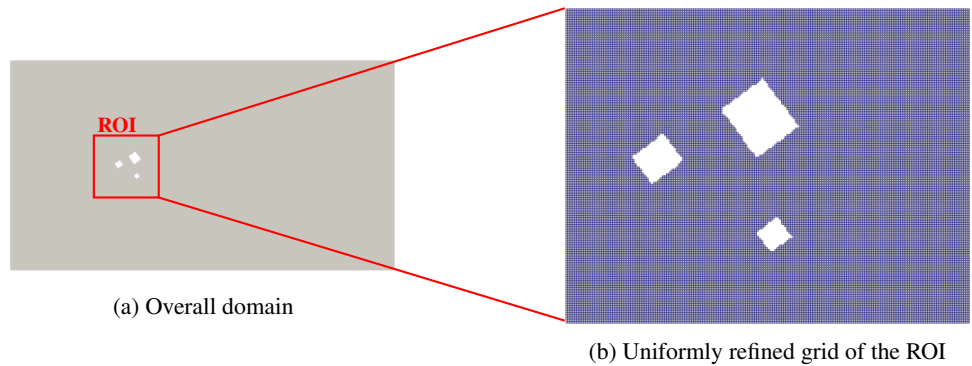
\begin{figure}[h]
    \centering
    \begin{tikzpicture}[remember picture]
        
        \node[inner sep=0] (main_node) at (0,0) {
            \begin{subfigure}[b]{0.3\textwidth}
                \centering
                \includegraphics[width=\textwidth]{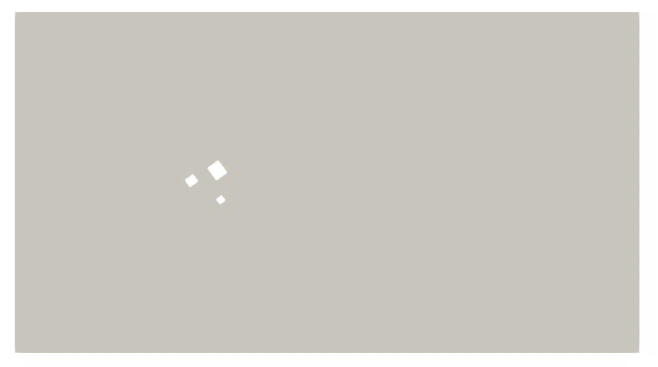}
                \caption{Overall domain}
                \label{fig:grid_0}
            \end{subfigure}
        };

        \node[inner sep=0] (zoom_node) at (7.5,0) {
            \begin{subfigure}[b]{0.3\textwidth}
                \centering
                \includegraphics[width=\textwidth]{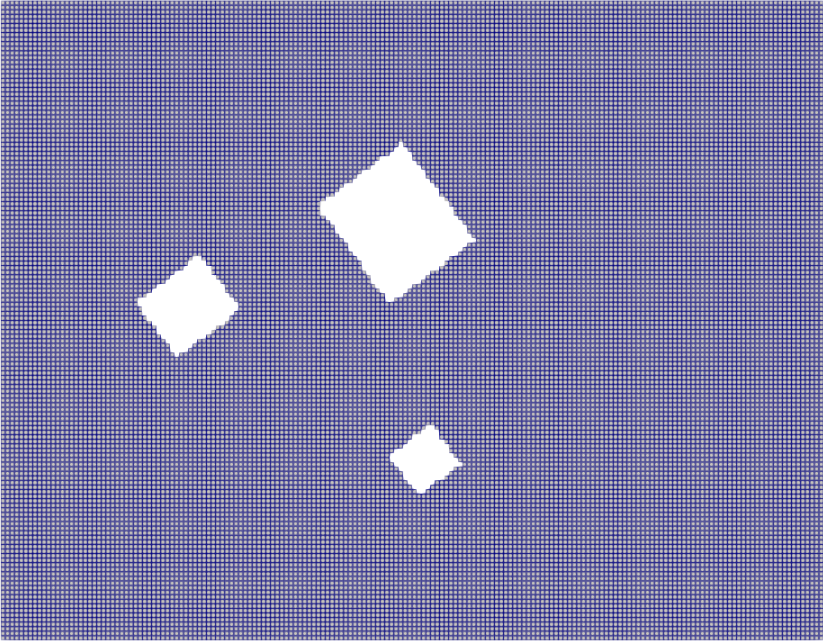}
                \caption{Uniformly refined grid of the ROI}
                \label{fig:grid_0_zoom}
            \end{subfigure}
        };

        \begin{scope}[shift={(main_node.south west)}, 
                      x={(main_node.south east)}, 
                      y={(main_node.north west)}, 
                      red, thick]
            \draw (0.23, 0.475) rectangle (0.39, 0.7) coordinate (roi1_top) at (0.39, 0.7) 
                                                     coordinate (roi1_bottom) at (0.39, 0.475);
            \node[anchor=south west, font=\bfseries\footnotesize, inner sep=1pt] at (0.23, 0.7) {ROI};

            \draw (roi1_top) -- (zoom_node.north west);
            \draw (roi1_bottom) -- ([yshift=2.2em]zoom_node.south west);
        \end{scope}

    \end{tikzpicture}
    
    \caption{Example for a uniform grid of the \ac{ROI}.}
    \label{fig:mesh_for_ID0}
\end{figure}

To facilitate the training of \ac{CNN} and \ac{GNN} models, 
the proposed dataset provides utilities for transforming the raw simulation output into \ac{ML}-compatible representations. 
As a basis, a uniformly refined, Cartesian grid covering the \ac{ROI} introduced in Sec.~\ref{subsec2} is provided.
The spatial dimensions of the \ac{ROI} are defined as $9d_{ref}$ in the x-direction and $7d_{ref}$ in the y-direction. 
Figure~\ref{fig:grid_0} shows the \ac{ROI} and a uniformly refined grid for the same example that was illustrated in Fig.~\ref{fig:mesh_refinement_levels}. 
Notably, this uniform grid is designed to be coarser than the original grid to minimize computational costs and enhance training efficiency for \ac{ML} models.

Two types of data mappings are considered.
For the first type, all flow variables are resampled from the original hierarchical grid onto the uniformly refined grid.
The mapping is performed using a nearest-neighbor assignment. 
For each cell of the target uniform grid, the value of the closest cell center in the original hierarchical grid is selected, avoiding additional smoothing or interpolation-induced artifacts while ensuring a one-to-one and computationally efficient transfer of flow variables.
This results in dense, regularly spaced arrays that are directly compatible with \ac{CNN}-based architectures and other grid-based learning methods. 
The mapping ensures a consistent spatial discretization across all samples, enabling straightforward batching, normalization, and comparison between different simulations.

For the second mapping type, the uniformly refined grid is used as the basis for constructing graph-based representations of the flow field. 
Each cell of the uniform grid is interpreted as a node in a graph, 
with edges defined according to spatial adjacency. 
From this representation, adjacency lists, edge indices, and optional edge attributes, such as relative position vectors or distances, are generated. 
The resulting graph structure, together with node-wise flow variables, serves as direct input for \ac{GNN}-based models while retaining a clear geometric interpretation.

To prepare the training data for \acp{CNN} and \acp{GNN}, 
data-handling scripts for both types of mapping are provided.
Details about the scripts are given in Sec.~\ref{subsec7}.
These preprocessing scripts 
enable the use of a wide range of learning architectures without requiring modifications to the underlying \ac{CFD} solver or storage format. At the same time, the original high-fidelity simulation data remains fully accessible, allowing users to tailor the resampling resolution or graph connectivity to the requirements of their specific application.


\section{Result}
\label{sec3}

This section analyzes simulation results that are either used to verify the dataset or belong to the dataset. 
In Sec.~\ref{subsec4}, grid refinement studies are conducted for a simplified case with flow around a single square cylinder and for a sample from the dataset with multiple buildings to determine a suitable grid resolution. 
In Sec.~\ref{subsec5}, simulation results using the grid resolution selected in Sec.~\ref{subsec4} are validated through a comparison of the single-cylinder flow with reference data from literature.
Section~\ref{subsec6} shows the flow fields of representative samples of the dataset at different \textsc{Reynolds} numbers,
and Sec.~\ref{subsec7} provides information about the data structure in the repository of the dataset.

\subsection{Grid Refinement Studies}
\label{subsec4}

Grid refinement studies are conducted to assess the influence of spatial resolution on the predicted flow fields and to determine a suitable grid configuration for the simulations included in the dataset. 
Ensuring sufficient grid resolution is essential for capturing key flow features such as wake structures and recirculation zones while maintaining computational efficiency across the large number of simulations. 
To this end, refinement analyses are performed for two representative configurations: a simplified benchmark case with flow around a single square cylinder and a multi-building configuration representative of the urban layouts contained in the dataset. 
The results of these studies are used to select an appropriate grid resolution for all subsequent simulations.

\subsubsection{Single-Building Case}
\label{subsubsec1}
\begin{figure}[h]
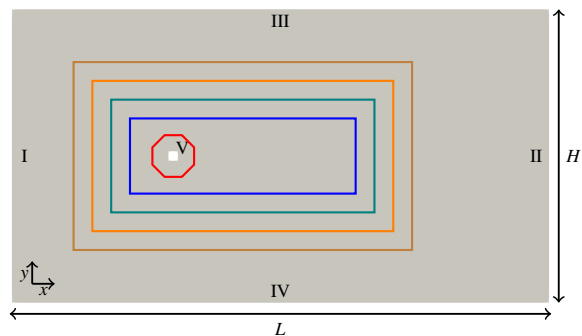

\centering
\include{figures/domain_grid_ref_one_column}
\caption{Simulation domain and local refinement regions for the grid refinement with flow around a single square cylinder.}
\label{fig:grid_ref_domain}
\end{figure}

The computational domain for the single-building case and various local refinement regions are shown in Fig.~\ref{fig:grid_ref_domain}. 
The building is represented by a non-rotated square cylinder with an edge length of $d_{ref}$.
The domain size and boundary conditions match with those described in Sec.~\ref{sec2}. 
The cell sizes and spatial extents (minimum and maximum coordinates) for the various refinement regions are listed in Tab.~\ref{tab:ref_patches}. 
The center of the square cylinder coincides with the origin of the coordinate system. 
The grid refinement study is performed at $Re = 500$, 
since no reliable reference data for 2D flow around a square cylinder at higher \textsc{Reynolds} numbers can be found in existing literature.
The reference data is required for the validation in Sec.~\ref{subsec5}.

\begin{table}[h]
\centering
\caption{Cell sizes and minimum and maximum coordinates for each refinement region for the coarse, medium and fine grids.}
\label{tab:ref_patches}
\footnotesize 
\begin{tabular}{|c|c|c|c|c|c|c|}
\hline
Region & $\Delta x_{coarse}$ & $\Delta x_{medium}$  & $\Delta x_{fine}$ & $|x_{min}|$, $|y_{min}|$, $y_{max}$ & $x_{max}$ \\
\hline
brown & $d_{ref}/3.125$ & $d_{ref}/6.25$ & $d_{ref}/12.5$ & $6d_{ref}$ & $24d_{ref}$ \\
\hline
orange & $d_{ref}/6.25$ & $d_{ref}/12.5$ & $d_{ref}/25$ & $5d_{ref}$ & $20d_{ref}$ \\
\hline
teal & $d_{ref}/12.5$ & $d_{ref}/25$ & $d_{ref}/50$ & $4d_{ref}$ & $16d_{ref}$ \\
\hline
blue & $d_{ref}/25$ & $d_{ref}/50$ & $d_{ref}/100$ &  $3d_{ref}$ & $12d_{ref}$ \\
\hline
red & $d_{ref}/50$ & $d_{ref}/100$ & $d_{ref}/200$ &  $0.64d_{ref}$ & $0.64d_{ref}$ \\
\hline
\end{tabular}
\end{table}

Table~\ref{tab:grid_ref} summarizes the minimum cell size $\Delta x_{min}$, total cell count $N_{cells}$, and the drag coefficient defined by
\begin{equation}
	C_{d, sq}=\frac{2 \cdot F}{\rho \cdot U_{ref}^2 \cdot d_{ref}},
\label{eq:Cd_sq}
\end{equation}

where $F$ is the drag force acting in the positive $x$-direction. 
The relative error of the drag coefficient between the medium and coarse grids is $+5.1\%$, while the error between the fine and medium grids is $+3.4\%$.
This decreasing trend in the relative error indicates quantitative grid convergence.

\begin{table}[h]
\centering
\caption{Minimum cell size, number of cells, and drag coefficient for flow around the single square cylinder at $Re=500$ with the coarse, medium, and fine grids. The percentages in parentheses indicate the relative deviation of the drag coefficient compared to the next coarser grid.}
\label{tab:grid_ref}
\footnotesize 
\begin{tabular}{|c|c|c|c|}
\hline
Grid & $\Delta x_{min}$ & $N_{cells} [\cdot 10^3]$ & $C_{d, sq}$ \\ \hline
Coarse & $d_{ref}/50$ & 120 & 1.72 (--) \\ \hline
Medium & $d_{ref}/100$ & 480 & 1.79 ($+5.1\%$) \\ \hline
Fine & $d_{ref}/200$ & 2,000 & 1.85 ($+3.4\%$) \\ \hline
\end{tabular}
\end{table}

\begin{figure}[h]
    \centering
    \makebox[\textwidth][c]{
    \begin{subfigure}[b]{0.35\textwidth}
        \centering
        \includegraphics[width=\textwidth]{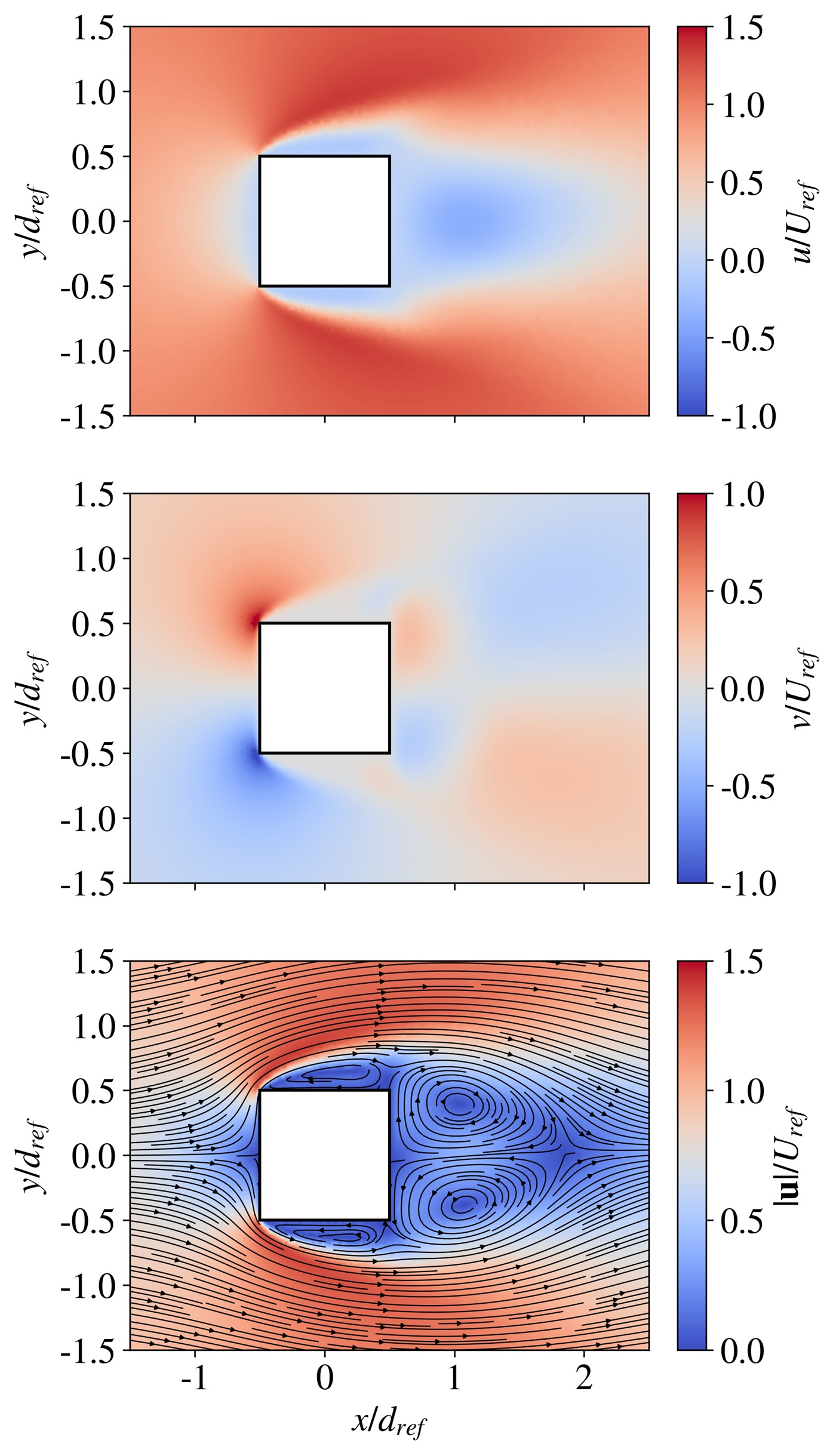} 
        \caption{Coarse grids}
        \label{fig:Re500_coarse}
    \end{subfigure}
    \hfill
    \begin{subfigure}[b]{0.35\textwidth}
        \centering
        \includegraphics[width=\textwidth]{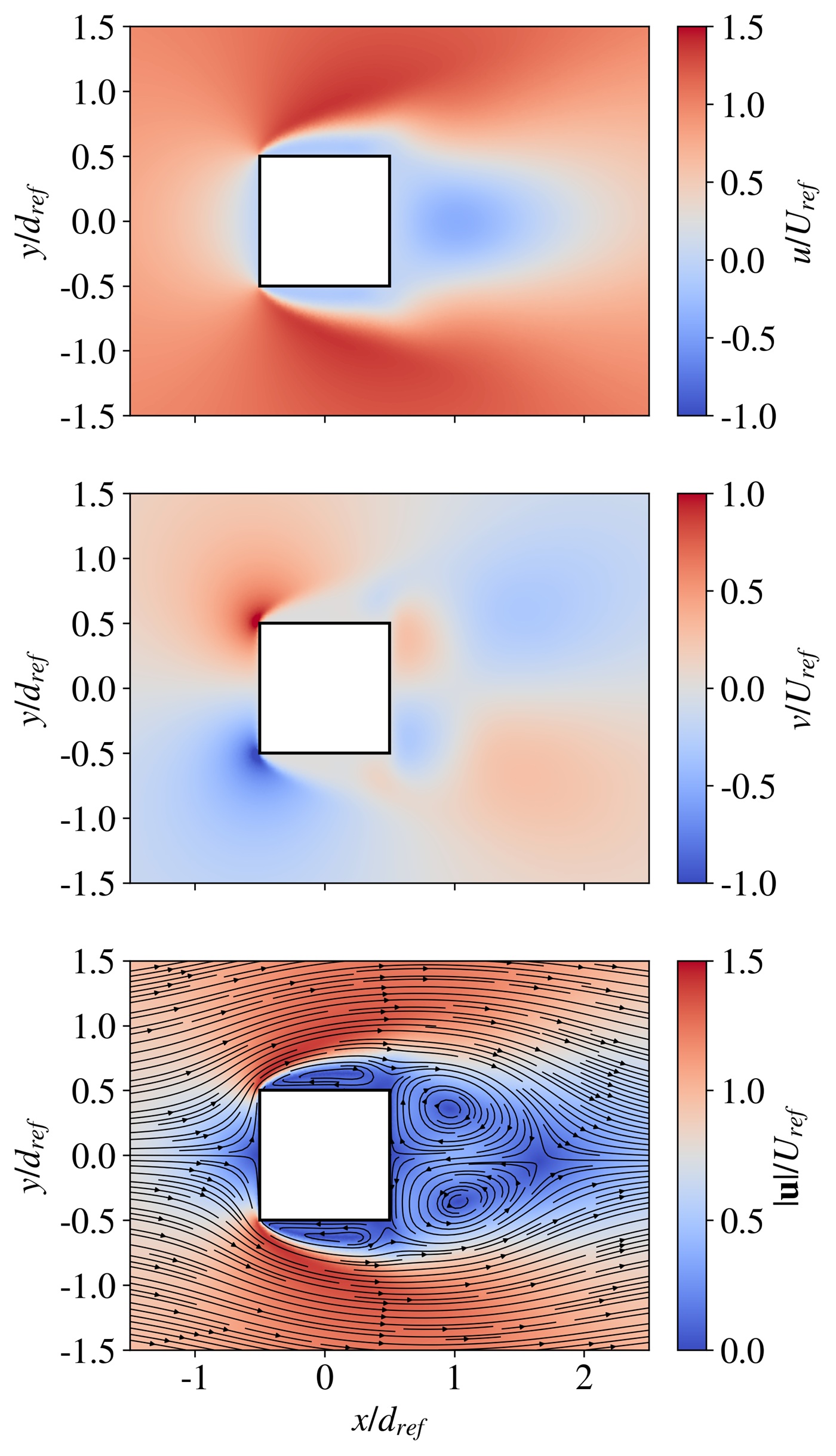}
        \caption{Medium grids}
        \label{fig:Re500_mid}
    \end{subfigure}
    \hfill
    \begin{subfigure}[b]{0.35\textwidth}
        \centering
        \includegraphics[width=\textwidth]{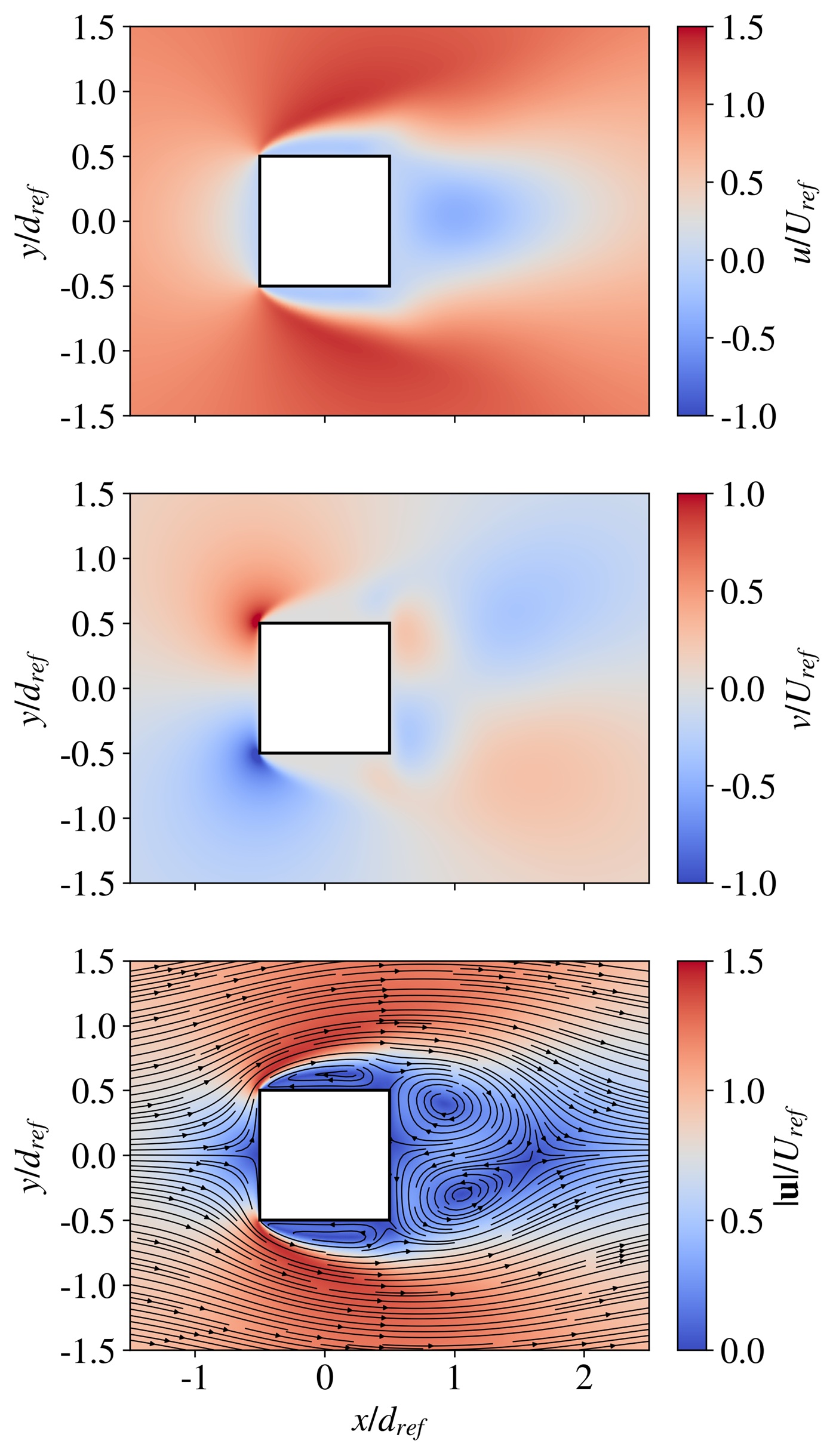}
        \caption{Fine grids}
        \label{fig:Re500_fine}
    \end{subfigure}
    }
    \caption{Time-averaged $u/U_{ref}$ (top row) and $v/U_{ref}$ (middle row) velocity fields and streamlines with $|\mathbf{u}|/U_{ref}$ (bottom row) for the three grid resolutions at $Re=500$.}
    \label{fig:full_results}
\end{figure}

Figure~\ref{fig:full_results} illustrates the time-averaged flow fields and streamlines around the single building for the three different grid resolutions. 
From top to bottom, the figures show the normalized $x$-component of the velocity vector ($u/U_{ref}$), the normalized $y$-component of the velocity vector ($v/U_{ref}$), and the normalized velocity magnitude $|\textbf{u}|/U_{ref}$ overlaid with streamlines, where $|\textbf{u}|$ denotes the velocity magnitude. 
In all cases, the $u/U_{ref}$ field clearly identifies an expected stagnation point at the front of the building along the centerline at $y=0$, followed by flow acceleration as it passes the corners. 
The $v/U_{ref}$ component exhibits its peak values near the corners of the front edges. 
The streamlines further reveal the formation of recirculation zones behind the building.
No visible differences between the results for the three grid resolutions support qualitative convergence with respect to the grid resolution.


\begin{figure}[htbp]
    \centering
    \makebox[\textwidth][c]{
    \begin{subfigure}[b]{0.55\textwidth}
        \centering
        \includegraphics[width=\textwidth]{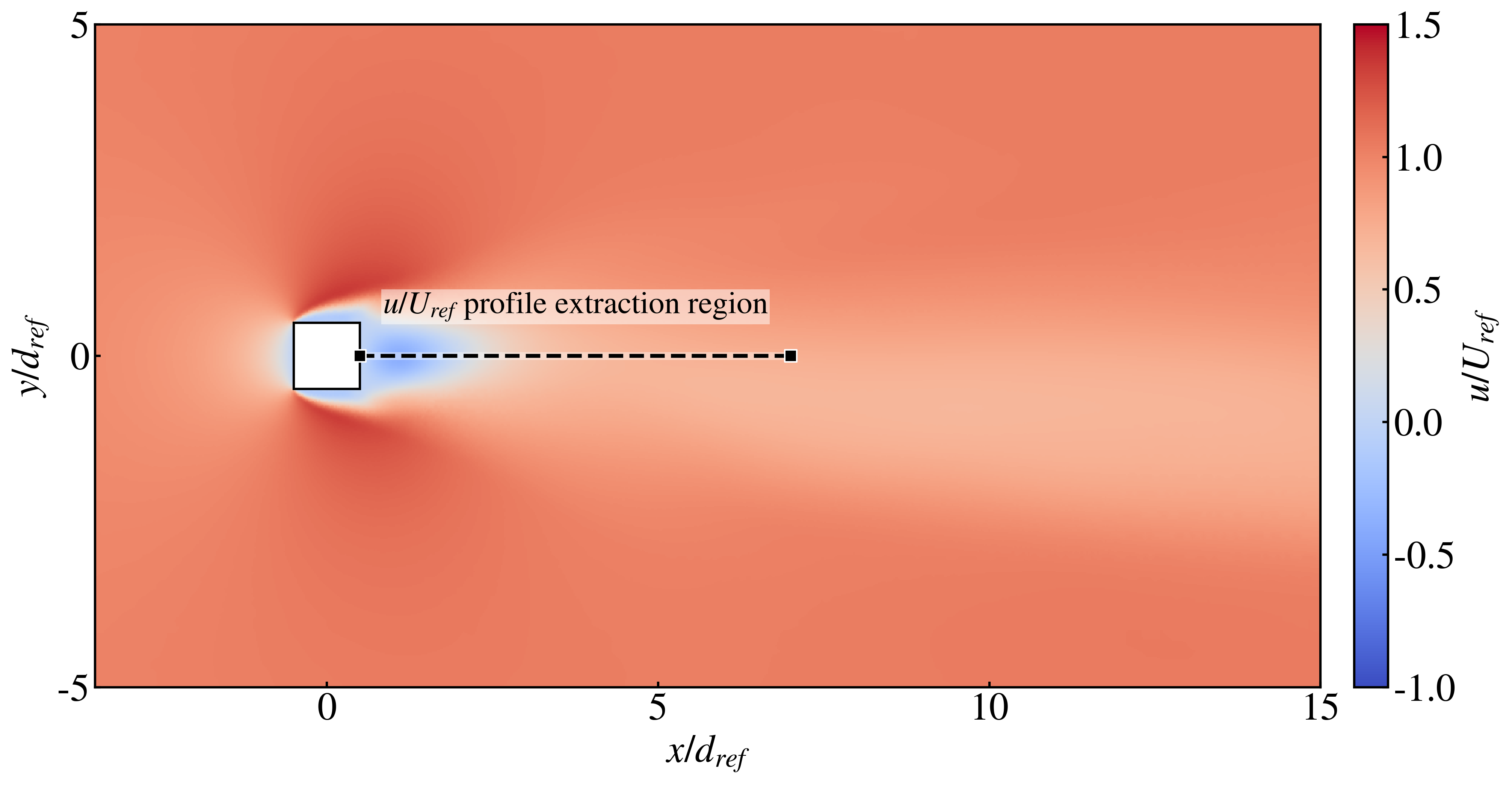}
        \caption{Time-averaged $u/U_{ref}$ flow field highlighting the spatial extend of the profiles shown in Fig.~\ref{fig:Re500_1Dplot} at $0.5 \leq x/d_{ref} \leq 7.0$ and $y/d_{ref}=0$.}
        \label{fig:Re500_dashed_line}
    \end{subfigure}
    \begin{subfigure}[b]{0.4\textwidth}
        \centering
        \includegraphics[width=\textwidth]{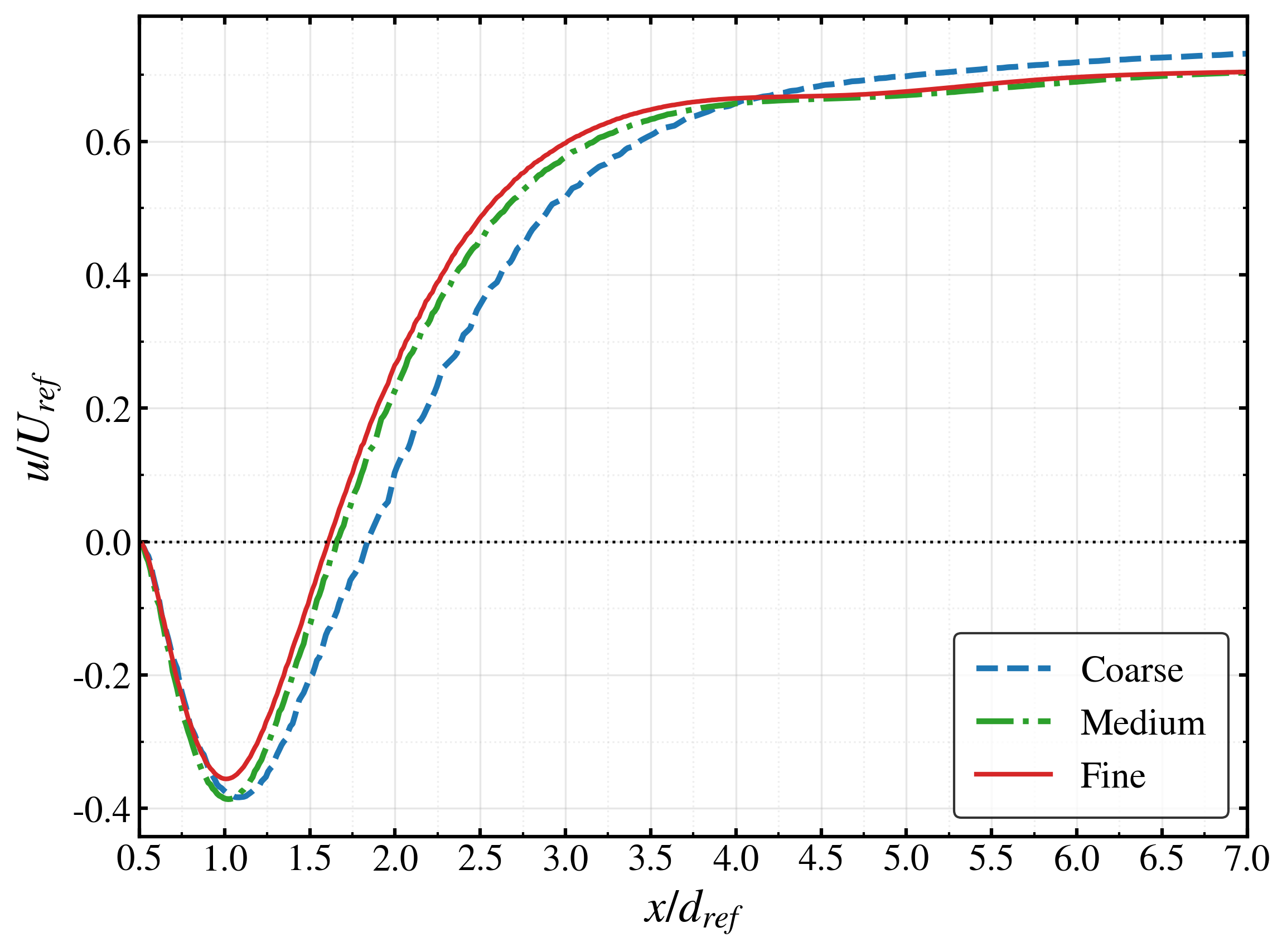} 
        \caption{Profiles of the time-averaged $u/U_{ref}$ along the dashed line in Fig.~\ref{fig:Re500_dashed_line}.}
        \label{fig:Re500_1Dplot}
    \end{subfigure}
    }
    \caption{Velocity profiles for flow around a single square cylinder for the three grid resolutions at $Re=500$.}
    \label{Re500_velocity_profile}
\end{figure}

Figure~\ref{Re500_velocity_profile} shows profiles of the time-averaged $u/U_{ref}$ for the three grids in the wake region of the square cylinder.
The dashed line in Fig.~\ref{fig:Re500_dashed_line} highlights the spatial extent of the profiles along the centerline $y/d_{ref} = 0$ plotted in
Fig.~\ref{fig:Re500_1Dplot}. 
The data are extracted from $x/d_{ref} = 0.5$, immediately downstream of the building, to $x/d_{ref} = 7.0$, ensuring a sufficient distance to observe the wake development. 
Due to the low-pressure zone behind the building, a recirculation region is formed, characterized by steep velocity gradients. 
Whereas the results with the coarse and medium grids deviate in the flow recovery and far wake regions for $x/d_{ref} \geq 1.0$,
the results with the medium and fine grids are nearly aligned.
The mean error along the velocity profile between the coarse and the medium grids, normalized by the mean velocity of the medium grid, is $5.3\%$,
while the mean error between the medium and fine grids, normalized by the mean velocity of the fine grid, is only $1.6\%$. 
Together with the drag coefficients and the flow fields shown in Fig.~\ref{fig:full_results}, this indicates that the results are qualitatively and quantitatively converged with respect to the grid resolution.

\subsubsection{Multi-Building Case}
\label{subsubsec2}
\begin{figure}[h]
    \centering
    \makebox[\textwidth][c]{
    \begin{subfigure}[b]{0.38\textwidth}
        \centering
        \includegraphics[width=\textwidth]{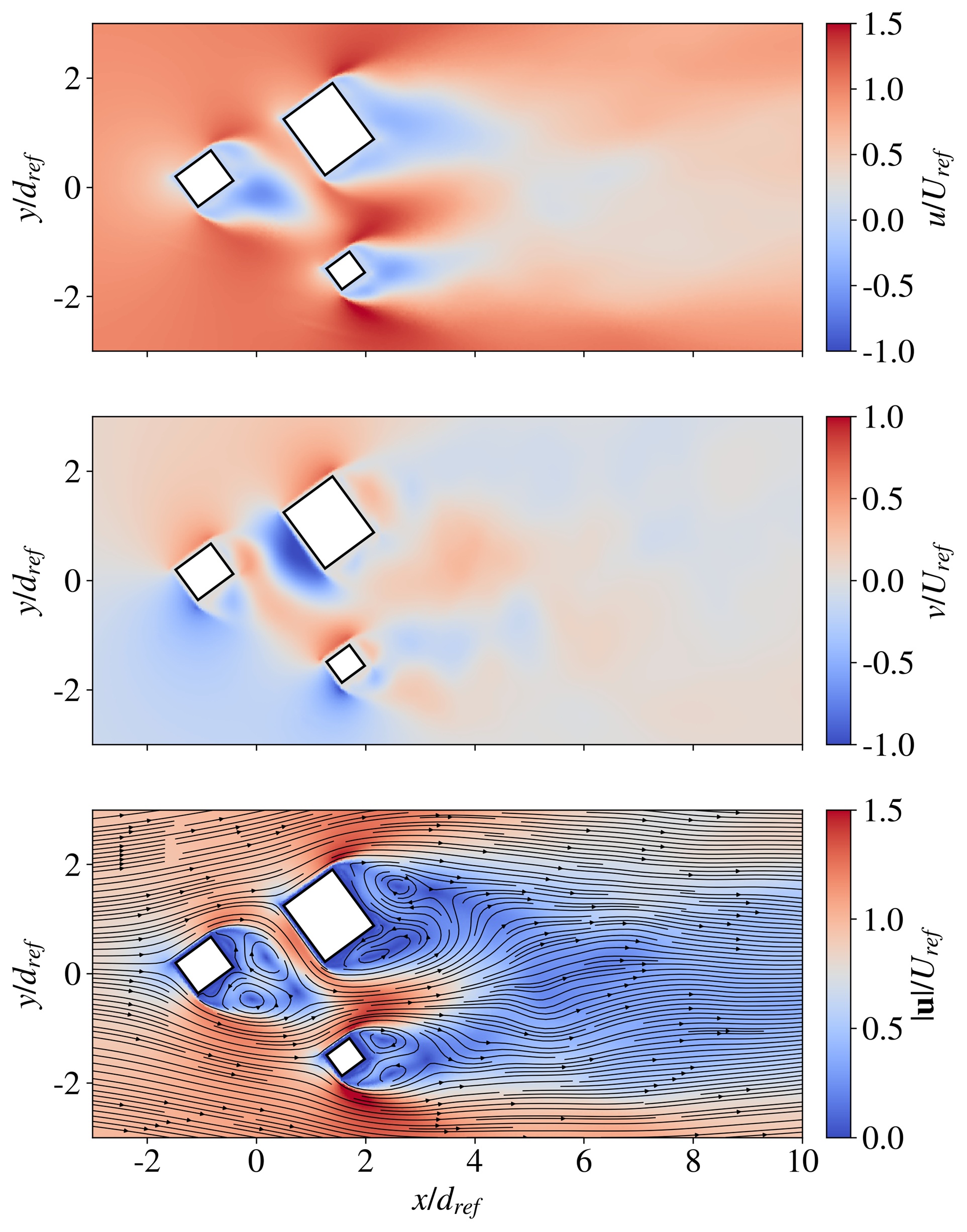} 
        \caption{Coarse grids}
        \label{fig:0_coarse}
    \end{subfigure}
    \hfill
    \begin{subfigure}[b]{0.38\textwidth}
        \centering
        \includegraphics[width=\textwidth]{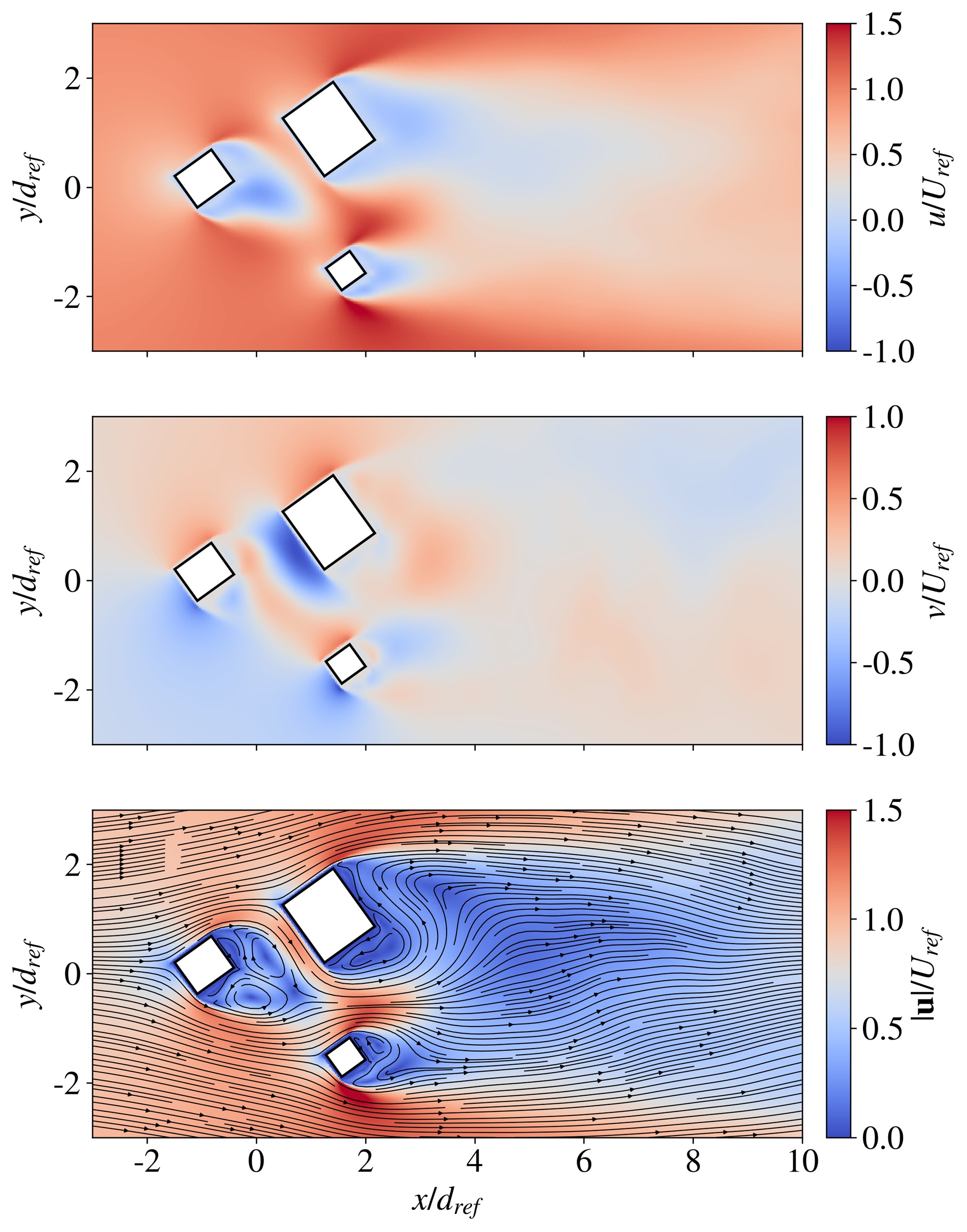}
        \caption{Medium grids}
        \label{fig:0_mid}
    \end{subfigure}
    \hfill
    \begin{subfigure}[b]{0.38\textwidth}
        \centering
        \includegraphics[width=\textwidth]{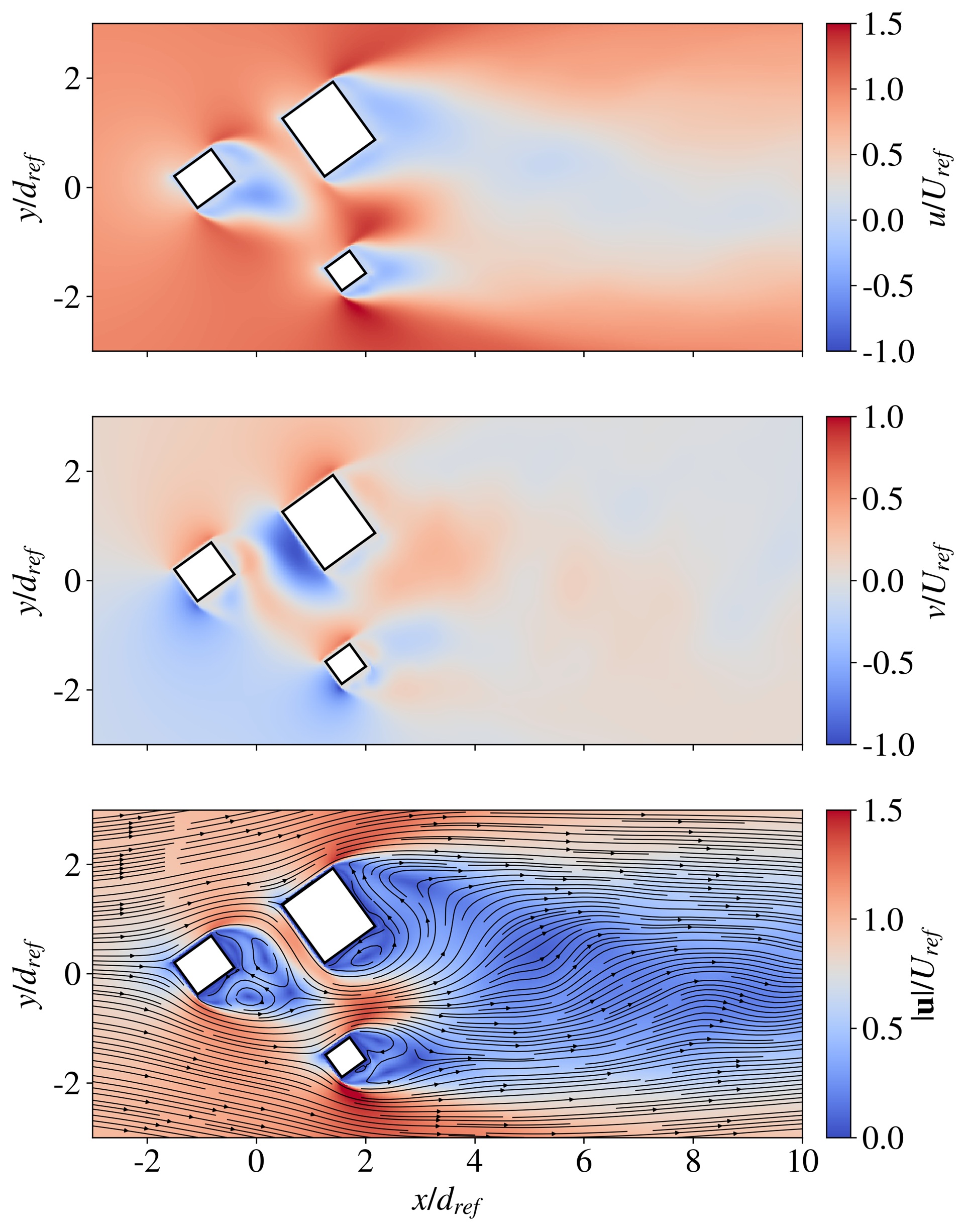}
        \caption{Fine grids}
        \label{fig:0_fine}
    \end{subfigure}
    }
    \caption{Time-averaged $u/U_{ref}$ (top row), $v/U_{ref}$ (middle row) velocity fields and streamlines with $|\mathbf{u}|/U_{ref}$ (bottom row) for the three grid resolutions at $Re=3{,}000$.}
    \label{fig:0_results}
\end{figure}

Consistent with the grid refinement study conducted in Sec. ~\ref{subsubsec1}, the analysis is extended from a single-building to a multi-building configuration from the dataset at $Re=3{,}000$.
Figure~\ref{fig:0_results} presents the time-averaged flow fields and streamlines around the multi-building arrangement for the three grid resolutions. 
The rows represent $u/U_{ref}$, $v/U_{ref}$, and $|\mathbf{u}|/U_{ref}$ with streamlines, respectively. 
In all cases, the stagnation points in the $u/U_{ref}$ flow fields consistently appear on the upper-left windward corners of the buildings, 
assuming a fixed orientation without rotation. 
The $v/U_{ref}$ component exhibits peak values at the top and bottom corners after rotation. 
Furthermore, the streamlines clearly capture the recirculation zones downstream of the buildings. 
The results from the coarse grid show a smaller extent of the wake region downstream of the largest building, a larger extent of the wake region downstream of the smallest building, and a larger jet between the two buildings, 
compared to the results obtained with the medium and fine grids.
This indicates qualitative convergence.

\begin{figure}[htbp]
    \centering
    \makebox[\textwidth][c]{
    \begin{subfigure}[b]{0.65\textwidth}
        \centering
        \includegraphics[width=\textwidth]{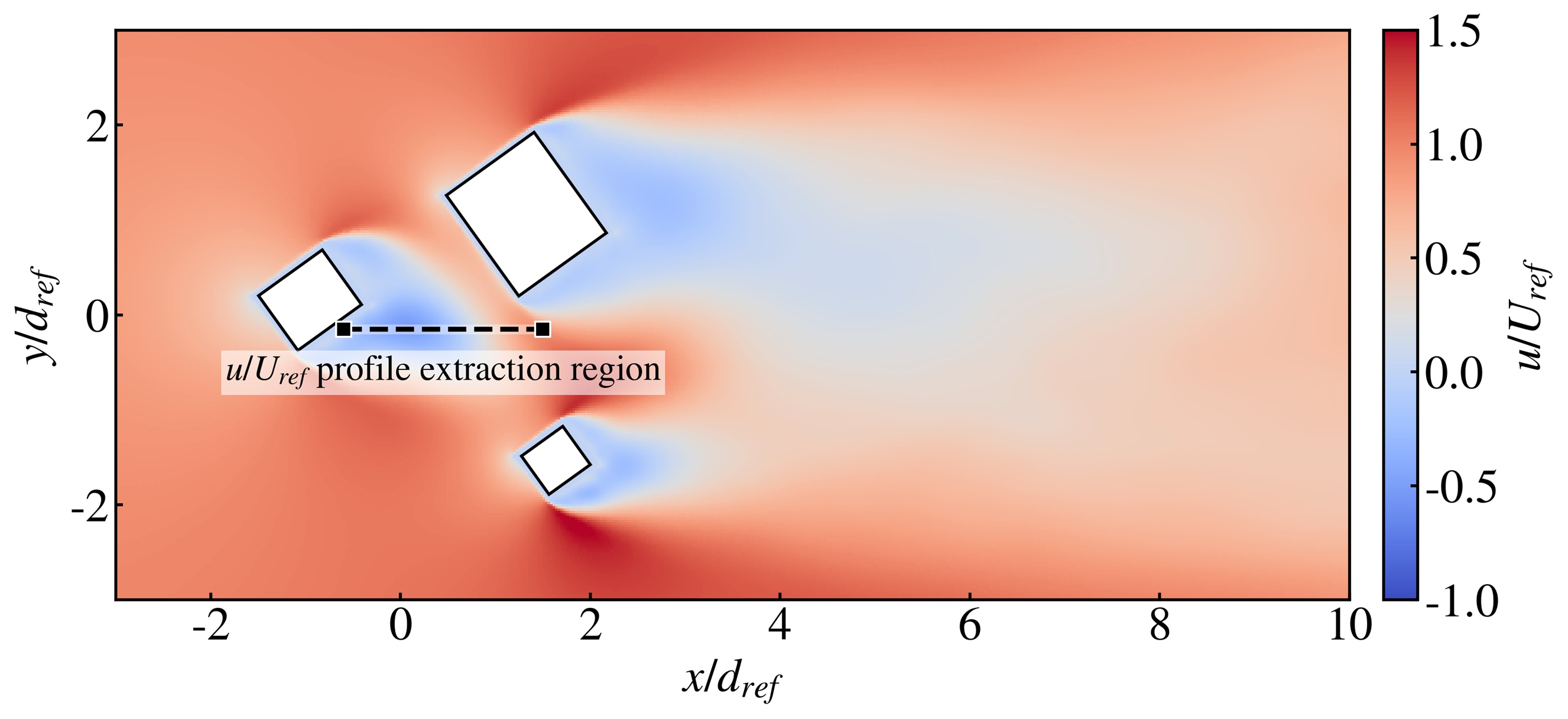}
        \caption{Time-averaged $u/U_{ref}$ flow field highlighting the spatial extend of the profiles shown in Fig.~\ref{fig:Re3000_1Dplot} at $-0.6 \leq x/d_{ref} \leq 1.5$ and $y/d_{ref}=-0.15$.}
        \label{fig:Re3000_dashed_line}
    \end{subfigure}
    \begin{subfigure}[b]{0.4\textwidth}
        \centering
        \includegraphics[width=\textwidth]{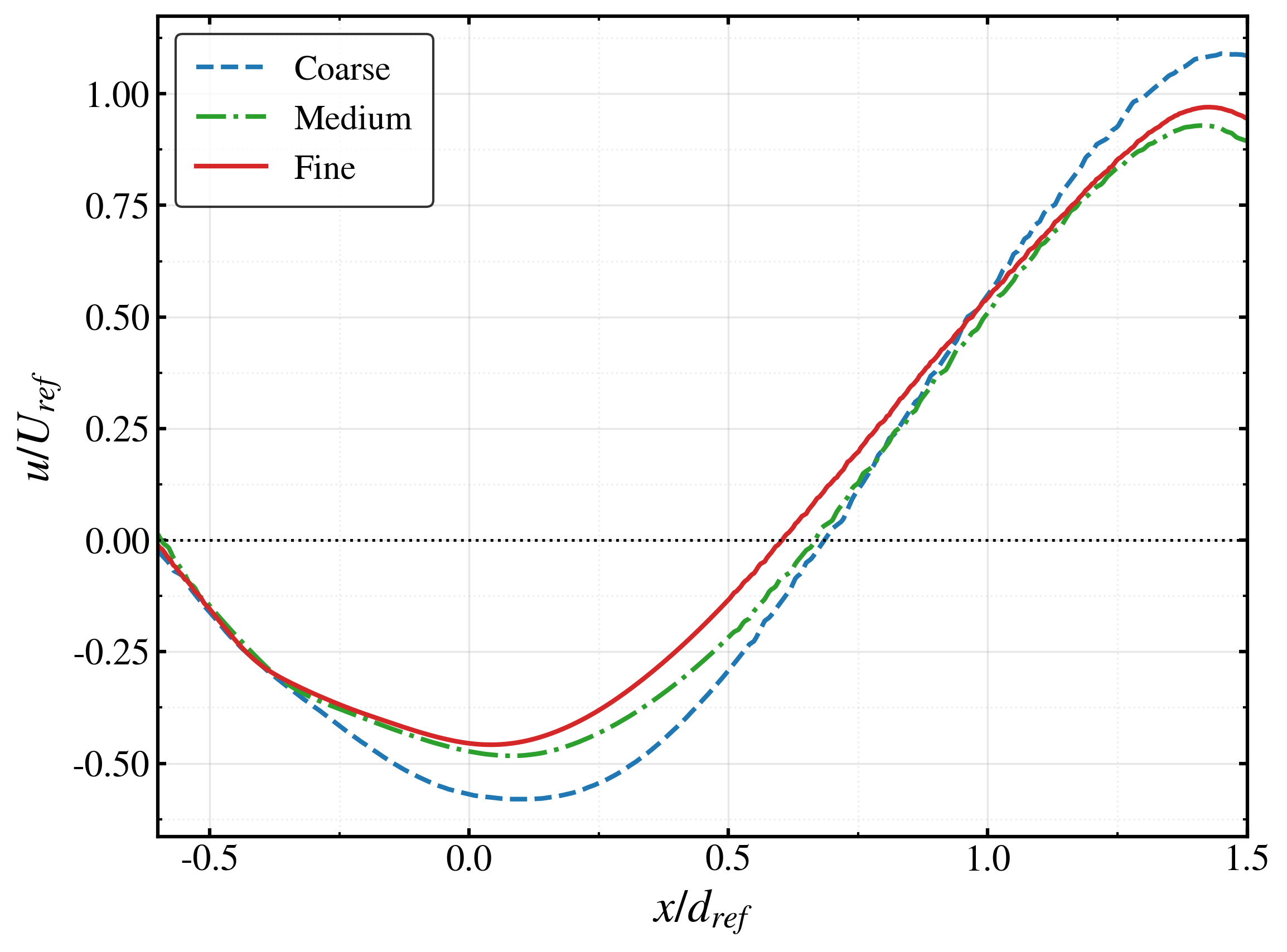} 
        \caption{Profiles of the time-averaged $u/U_{ref}$ along the dashed line in Fig.~\ref{fig:Re3000_dashed_line}}
        \label{fig:Re3000_1Dplot}
    \end{subfigure}
    }
    \caption{Velocity profiles for flow around multiple buildings at $Re=3{,}000$.}
    \label{0_velocity_profile}
\end{figure}

Figure~\ref{0_velocity_profile} shows time-averaged velocity profiles for the three grids in the wake region of the building with the center near $x/d_{ref}=-1$.
The dashed line in Fig.~\ref{fig:Re3000_dashed_line} illustrates the spatial extent of the $u/U_{ref}$ profiles plotted in
and Fig.~\ref{fig:Re3000_1Dplot}.
To encompass the minimum velocity region located near $x/d_{ref} = 0.05$ and $y/d_{ref} = -1.5$, the $u/U_{ref}$ profiles were extracted along the line $y/d_{ref} = -1.5$ within the range $-0.6 \leq x/d_{ref} \leq 1.5$ for the coarse, medium, and fine grids. 
While distinct discrepancies are observed between the coarse and fine grids, the profiles of the medium and fine grids are nearly aligned. 
The mean error between the coarse and medium grids, normalized by the mean velocity of the medium grid, is $8.7\%$. 
In contrast, the error between the medium and fine grids is significantly reduced to $3.8\%$. 
These quantitative results demonstrate grid convergence for the complex multi-building domain.

Overall, both the qualitative comparison of the flow fields and streamlines and the quantitative analysis of the velocity profiles and error metrics demonstrate grid convergence for the single- and multi-building configurations. 
In both cases, the medium and fine grids show close agreement, whereas the coarse grid exhibits noticeable deviations, particularly in the wake and recirculation regions. 
Since the discrepancies between the medium and fine grids are small while the computational cost of the fine grid is substantially higher, the medium grid resolution is selected for the validation study presented in the following section.

\subsection{Validation}
\label{subsec5}
For the selected medium grid, a validation is performed through a comparison of the resulting drag coefficient, \textsc{Strouhal} number, and streamline patterns with the reference data in~\cite{Sohankar1999}.

Comparing the drag coefficient listed in Tab.~\ref{tab:grid_ref} with the reference value $C_{d, sq}^{ref}=1.89$ yields a relative error of $5.3\%$. 
Thus, the medium grid captures the drag coefficient with reasonable accuracy. 

In addition to the drag coefficient, the vortex shedding behavior is assessed through the \textsc{Strouhal} number
\begin{equation}
	St=\frac{f_t d_{ref}}{U_{ref}},
\label{eq:St}
\end{equation}
where $f_t$ denotes the dominant shedding frequency.
This frequency is extracted from the temporal velocity signal recorded at a point probe located at $6d_{ref}$ downstream of the cylinder center along the wake centerline. 
After discarding the initial transient phase, the signal is detrended to remove mean and linear components. 
A fast Fourier transform is then applied to the fluctuating velocity signal, 
and the dominant shedding frequency is identified as the frequency corresponding to the maximum spectral amplitude. For the medium grid resolution, a \textsc{Strouhal} number of $St=0.14$ is obtained.
Based on values reported in the literature, $St$ for 2D flow around a single square cylinder at $Re=500$ is generally expected to lie in the range of $St\in[0.13,0.18]$~\cite{Sohankar1999}.
The present result falls well within this commonly reported interval and is therefore considered physically consistent. 


Given the reasonable drag coefficient and \textsc{Strouhal} number, the medium grid resolution is selected for the simulations of the proposed dataset.
The grids of the dataset have between $0.5\cdot10^6$ and $1.5\cdot10^6$ cells.
Four simulations are run simultaneously on one \ac{GPU} node for $6$-$8$ hours, i.e., one simulation is assigned to each \ac{GPU}.
The uniformly refined grids of the \ac{ROI} have a cell size of $d_{ref}/20$ and consist of approximately $22{,}000$ to $25{,}000$ cells. 
The number of cells and simulation run time vary slightly across cases depending on the number and size of the buildings.


\subsection{Flow fields of representative samples of the dataset at different \textsc{Reynolds} numbers}
\label{subsec6}

\begin{figure}
    \centering
    \makebox[\textwidth][c]{
        \begin{subfigure}[b]{1.15\textwidth}
            \centering
            \includegraphics[width=\textwidth]{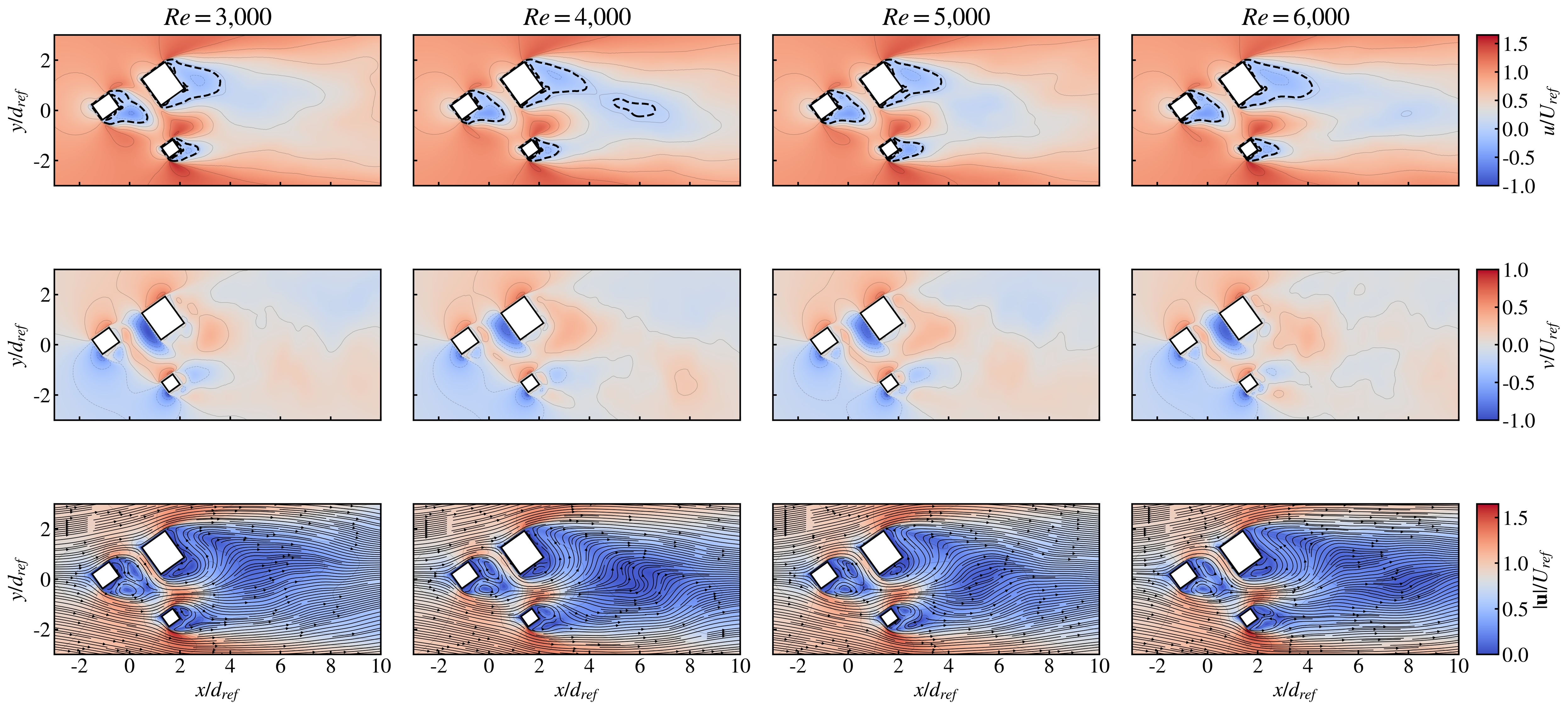}
            \label{fig:all_cases}
        \end{subfigure}
    }
    \caption{Time-averaged $u/U_{ref}$ (top row) and $v/U_{ref}$ (middle row) velocity fields and streamlines with $|\mathbf{u}|/U_{ref}$ (bottom row) for flow in the same multi-building configuration as in Secs.~\ref{subsec4} and~\ref{subsec5} at various \textsc{Reynolds} numbers $Re\in\{3{,}000, 4{,}000, 5{,}000, 6{,}000\}$.}
    \label{fig:Re3000_result_1}
\end{figure}

Figure~\ref{fig:Re3000_result_1} visualizes the time-averaged flow fields, streamlines, and the zero contour of $u/U_{ref}$ (dashed black line) for various \textsc{\textsc{Reynolds}} numbers using the same building configuration as in Fig.~\ref{fig:0_results}. 
The locations of flow separation at the building corners, which are identifiable in the $u/U_{ref}$ flow fields, 
as well as the peak velocities appearing above and below the building corners, which are visible in the $v/U_{ref}$ flow fields, 
remain nearly consistent across all \textsc{\textsc{Reynolds}} numbers. 
Comparing the flow fields at different \textsc{\textsc{Reynolds}} numbers, slight discrepancies in the zero-velocity contours and streamlines are observed.
For example, at $Re=4{,}000$ an additional recirculation zone appears in the far-wake region, which again disappears at $Re=5{,}000$.
However, increasing the \textsc{Reynolds} number further from $5{,}000$ to $6{,}000$ yields no significant alterations in the building wake or recirculation regions. 
This suggests that the $Re\in\{3{,}000, 4{,}000, 5{,}000\}$ cases are sufficient to generate a dataset with a rich variety of flow phenomena.

\begin{figure}[]
    \centering
    \makebox[\textwidth][c]{
        \begin{subfigure}[b]{0.37\textwidth}
            \centering
            \includegraphics[width=\textwidth]{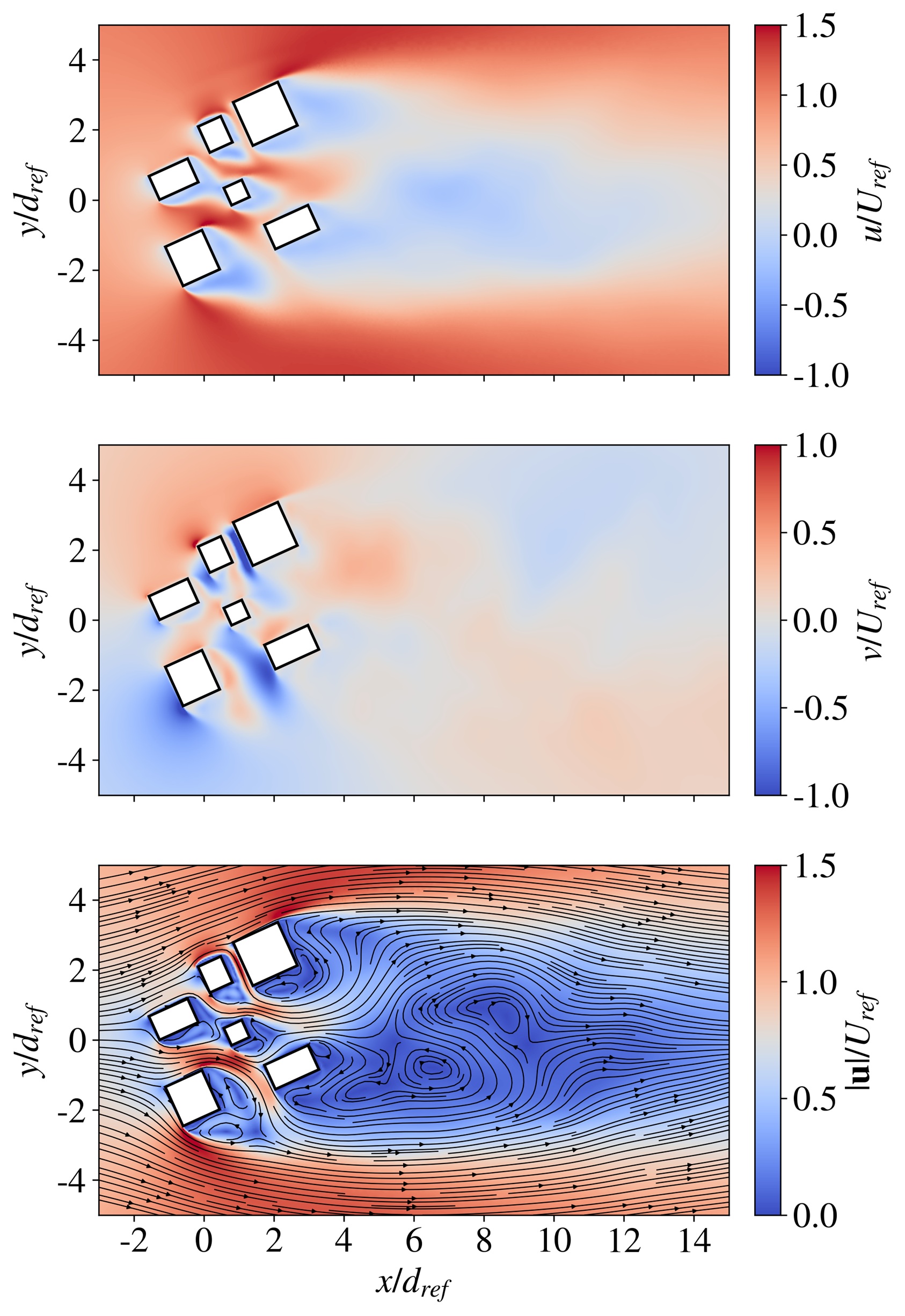}
            \caption{An example for $Re=3{,}000$}
            \label{fig:Re3000_ID3_uv}
        \end{subfigure}
        \hspace{0.5em}
        \begin{subfigure}[b]{0.37\textwidth}
            \centering
            \includegraphics[width=\textwidth]{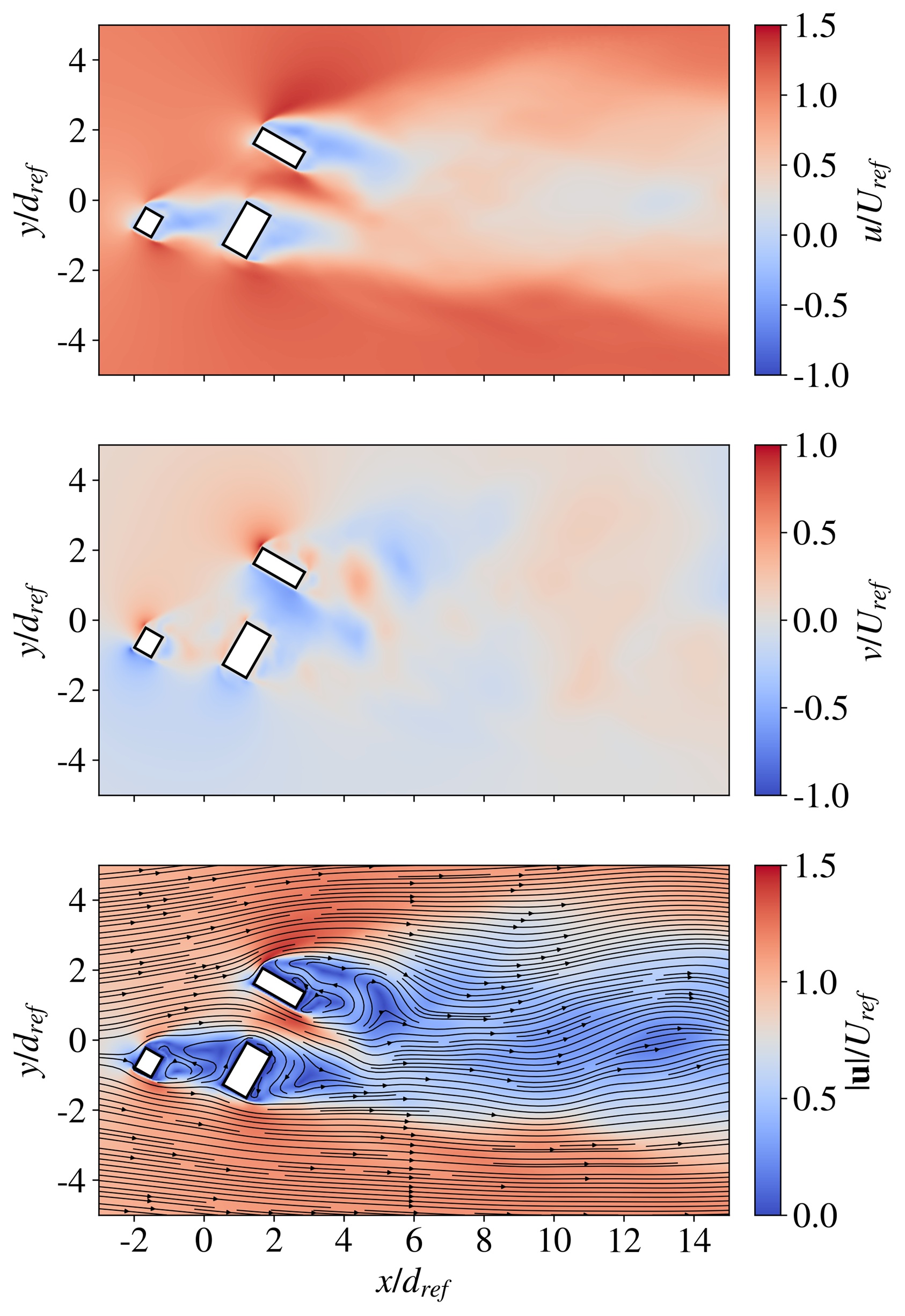}
            \caption{An example for $Re=4{,}000$}
            \label{fig:Re4000_ID1500_uv}
        \end{subfigure}
        \hspace{0.1em}
        \begin{subfigure}[b]{0.37\textwidth}
            \centering
            \includegraphics[width=\textwidth]{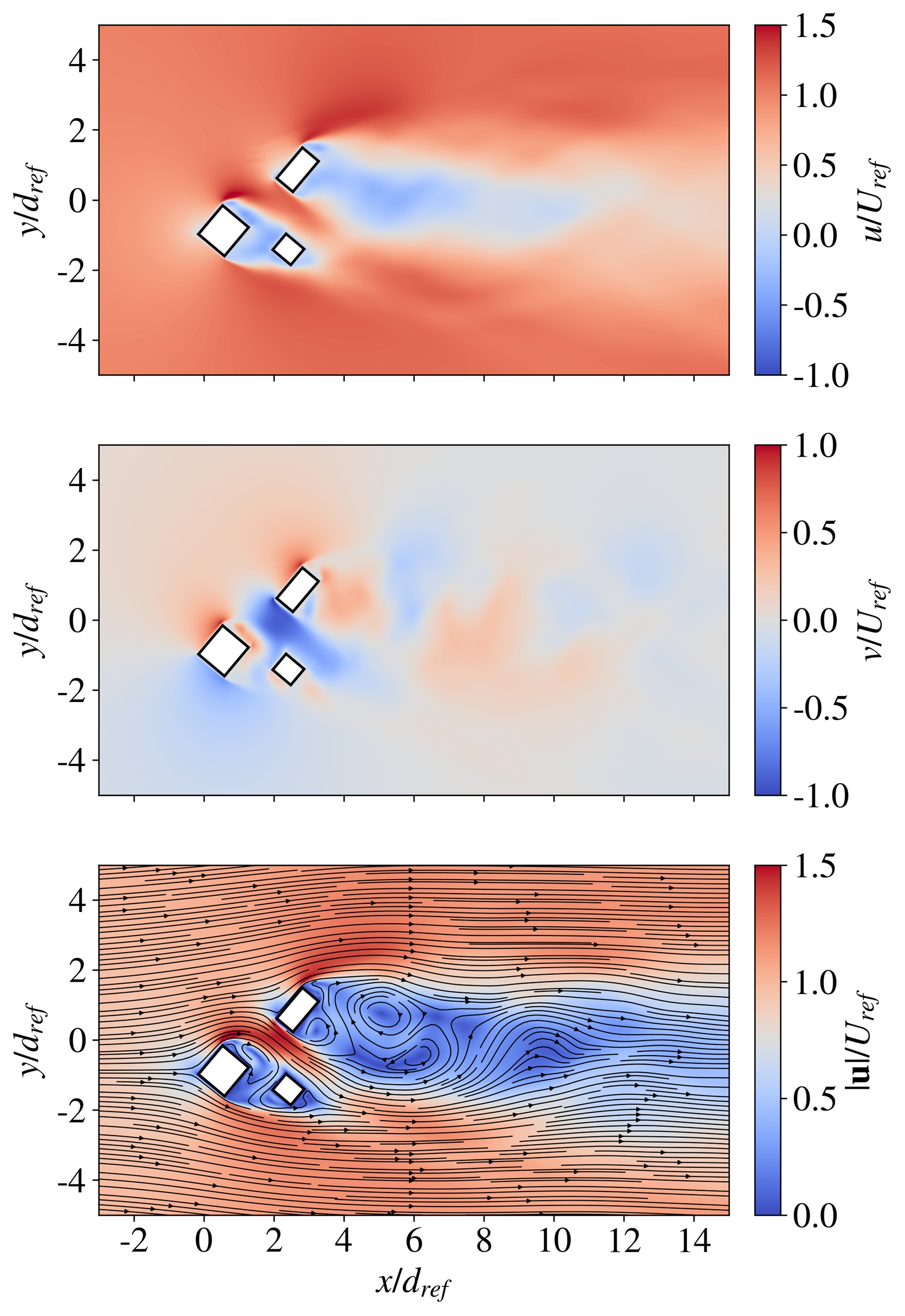}
            \caption{An example for $Re=5{,}000$}
            \label{fig:Re5000_ID2500_uv}
        \end{subfigure}
    }
    \caption{Examples of the time-averaged $u/U_{ref}$ (top row) and $v/U_{ref}$ (middle row) velocity fields and streamlines with $|\mathbf{u}|/U_{ref}$ (bottom row) for flow with various building configurations and $Re$ numbers.}
    \label{fig:Re3000_result_2}
\end{figure}

Figure~\ref{fig:Re3000_result_2} presents the flow fields and streamlines for different building configurations at these three \textsc{Reynolds} numbers. 
The configuration in Fig.~\ref{fig:Re3000_ID3_uv} contains six buildings. Due to the large number of buildings, the recirculation regions generated behind the buildings spread widely in the $y$-axis direction, accompanied by numerous vortical structures. 
In particular, a large-scale upward flow interacting with vortices at smaller scales is captured in the region of $3 \leq x/d_{ref} \leq 10$. 
For the upper-right building, the wake vortex formed downstream induces an upward deflection of the flow. In contrast, for the lower-right building, the corresponding vortex structure leads to a downward deflection of the flow.
Figure~\ref{fig:Re4000_ID1500_uv} shows a configuration at $Re=4{,}000$ with only three buildings.
Since the spacing between the buildings is larger compared to the previous configuration,
more separated recirulation regions and larger jets between buildings are observed.
Figure~\ref{fig:Re5000_ID2500_uv} shows a configuration at $Re=5{,}000$ with three buildings that are denser located compared to the case at $Re=4{,}000$. Flow separation occurs at the first building, and a strong jet forms between the buildings located downstream. 
leading to strong local flow acceleration. The uppermost building has a relatively larger aspect ratio in the flow direction compared to that in Fig.~\ref{fig:Re4000_ID1500_uv}, resulting in a wider recirculation region behind the building and a more clearly defined vortex structure. 
A large-scale upward flow interacting with the vortices generated in the range of $3 \leq x/d_{ref} \leq 7$ is also observed. 

Overall, the extent and interaction of the recirculation regions as well as the strength and confinement of the jets between buildings, strongly depend on the number of buildings and their spatial density. Configurations with fewer and more widely spaced buildings exhibit more isolated wake structures and distinct jet formations, whereas denser arrangements promote stronger flow channeling, enhanced jet acceleration, and increasingly interacting recirculation zones. This highlights the sensitivity of the wake dynamics to geometric complexity and building spacing.

\subsection{Dataset Structure}
\label{subsec7}
The dataset is hosted at the Urbanflow-3K repository with the DOI: \href{https://doi.org/10.7910/DVN/J2DRQO}{10.7910/DVN/J2DRQO}.
To guarantee a clear distinction between different simulations, 
an \ac{ID} is assigned to each simulation.
The \acp{ID} have the following order:
\begin{equation}
    \textnormal{ID}\in
    \begin{cases}
        \{0,\ldots,999\}, & Re=3{,}000\\
        \{1{,}000,\ldots,1{,}999\}, & Re=4{,}000\\
        \{2{,}000,\ldots,2{,}999\}, & Re=5{,}000
    \end{cases}.
\end{equation}
The dataset comprises three primary types of NetCDF-files, i.e., a grid, coordinate, and solution file.
The file names and key information for these three file types are as follows, where \texttt{ID} is a placeholder for the corresponding sample ID:

\begin{description}
    \item[\texttt{grid\_ID.Netcdf}] Contains the locally refined grid, including the cell ID, a unique identifier assigned to each individual grid cell, and the cell type, which defines the refinement level of a cell.
    \item[\texttt{Info\_ID.Netcdf}] Contains the $x$- and $y$-coordinates for the center of each cell of the locally refined grid, and the IDs of the neighboring cells in the positive and negative $x$- and $y$-directions.
    \item[\texttt{Mean\_s1\_2000000-2500000\_ID.Netcdf}] Contains flow field information, including time-averaged velocity components and fluid density for the locally refined grid.
    \item[\texttt{grid\_uniform\_ID.Netcdf}] Contains the uniformly refined grid of the \ac{ROI}, including the corresponding cell ID.
    \item[\texttt{Info\_uniform\_ID.Netcdf}] Contains the $x$- and $y$-coordinates and cell IDs of neighbouring cells for the uniformly refined grid within the \ac{ROI}
\end{description}

The first three NetCDF file types contain the same number of rows, corresponding to the total number of grid cells of a simulation. 
The row ordering is consistent across files and follows the Hilbert space-filling curve described in Sec.~\ref{sec2}, 
ensuring a one-to-one correspondence between grid topology, coordinates, and solution variables. 
Each row may contain multiple columns depending on the file type: for example, the \texttt{Info\_ID.Netcdf} file contains separate columns for the $x$- and $y$-coordinates of the cell centers, 
whereas the solution file contains one column per macroscopic variable, such as fluid density and the velocity components $u$ and $v$.
All simulations have been run for $2.5\cdot10^6$ time steps.
During the first $10^6$ time steps, the inflow velocity was gradually increased following a $\tanh$ curve to maintain numerical stability.
The range $[2\cdot10^6,2.5\cdot10^6]$ in case of the \texttt{Mean\_s1\_2000000-2500000\_ID.Netcdf} file represents the the time-averaging period, i.e., time-averaging was performed for $0.5\cdot10^6$ time steps.
This time-averaging interval was found to be necessary in a preliminary study to obtain statistically converged mean flow quantities. The files \texttt{grid\_uniform\_ID.Netcdf} and \texttt{Info\_uniform\_ID.Netcdf} provide grid information and cell coordinates for the uniformly refined grid of the \ac{ROI} to support \ac{CNN}- and \ac{GNN}-based learning.

Two data-handling scripts are provided to facilitate data preparation for \ac{CNN} and \ac{GNN} architectures:
\begin{description}
    \item[\texttt{gen\_CNN.ipynb}] Maps original results to the uniformly refined \ac{ROI} via a nearest-neighbor approach.
    \item[\texttt{gen\_GNN.ipynb}] Transforms the results from the uniformly refined \ac{ROI} into a graph-based representation of the flow field.
\end{description}

The execution of \texttt{gen\_CNN.ipynb} maps original outputs onto a uniformly refined \ac{ROI}, thereby generating structured datasets compatible with \acp{CNN}.

The execution of \texttt{gen\_GNN.ipynb} generates three distinct types of output files: \texttt{flow\_ID.hdf5}, which stores time-averaged velocity components and fluid density; \texttt{adjLst\_ID.hdf5}, which represents the topological connectivity and edge information between nodes; and \texttt{coord\_ID.hdf5}, which stores the node-level feature matrix including spatial coordinates.



\section{Discussion and conclusion}
\label{sec4}

High-fidelity \ac{CFD} simulations provide detailed insights into urban flow phenomena, but their substantial computational cost often limits their use in large parametric studies or real-time applications. 
As a consequence, ML-based surrogate models are increasingly being explored to accelerate urban flow field predictions. 
The development and validation of such models, however, critically depend on the availability of large, diverse, and well-structured training datasets. 
While several recent datasets provide highly resolved three-dimensional flow fields for realistic urban environments, 
their generation remains computationally expensive and therefore limits the number of available samples. 
Moreover, the complexity and computational cost associated with training on such datasets can make them less suitable for early-stage development and debugging of ML models, where rapid experimentation and large numbers of training samples are often required.

The dataset presented in this work complements existing three-dimensional urban flow datasets by providing a large collection of $3{,}000$ systematically generated two-dimensional urban flow simulations. 
The layouts incorporate between three and six buildings with randomized sizes, positions, and rotation angles, 
covering a wide range of obstacle interactions and effective wind directions. 
This controlled yet diverse configuration space enables the dataset to capture key flow features such as wake formation, shielding effects, flow acceleration, and recirculation zones across many urban layouts. 
At the same time, the relatively low computational cost of two-dimensional simulations allows the generation of substantially larger training sets than would typically be feasible with three-dimensional \ac{CFD} simulations.

Beyond providing a new benchmark dataset, the main motivation of this work is to support the systematic development and evaluation of data-driven flow prediction methods. 
The large sample size and consistent simulation setup enable controlled investigations of ML architectures, training strategies, and data requirements, which are often difficult to perform with computationally expensive three-dimensional datasets. 
In addition, the dataset can serve as a basis for transfer-learning approaches, in which models are first trained on large two-dimensional datasets before being adapted to smaller and more expensive three-dimensional urban flow datasets.

Alongside the comprehensive collection of $3{,}000$ simulations, two data-handling scripts designed to enhance accessibility for \ac{ML} training are also provided, 
thereby offering a more integrated and user-friendly dataset package. 
Users are encouraged to adapt these routines, 
for example, by employing alternative interpolation schemes or graph connectivity, 
depending on the target \ac{ML} model and research objective.

To ensure the accuracy of the generated dataset, a two-stage process consisting of a grid refinement study and validation was conducted. The grid refinement study was performed to determine a grid resolution that shows the best compromise between accuracy and computational costs for generating the dataset, evaluating the convergence of velocities and drag coefficients.
Results with the selected grid resolution were validated by a comparison of the drag coefficient and \textsc{Strouhal} number with the numerical study in~\cite{Sohankar1999}.

The dataset encompasses a diverse range of flow fields by combining various \textsc{Reynolds} numbers and building layouts. Notably, no significant structural changes were observed in the wake or recirculation regions as the \textsc{Reynolds} number increased from $Re=5{,}000$ to $Re=6{,}000$. 
This suggests that the dataset provides physically rich flow phenomena that remain valid and representative even at $Re\in\{3{,}000, 4{,}000, 5{,}000\}$. 
A consistent physical phenomenon observed across the results was the significant dependency of the wake structure on the building arrangement. Specifically, higher building density lead to jet acceleration between structures and the merging of individual wakes, resulting in vigorous interactions within the recirculation zones. 
In contrast, lower building density allowed for the formation of more isolated wakes. 
These observations are consistent with the research by Islam et al.~\cite{Islam2017}, which states that the arrangement and spacing of bluff bodies dictate the flow regime, i.e., larger spacing reduces wake interaction, while denser arrangements lead to stronger interference. 

Building on this dataset, 
ongoing work aims to extend the present data toward a substantially larger three-dimensional urban flow dataset using the \textit{urbanFlowGen} library (available at \url{https://github.com/MarioRuettgers/urbanFlowGen}). 
The forthcoming dataset will include both systematically generated synthetic urban layouts and samples derived from realistic city geometries, 
thereby combining controlled variability with real-world complexity. 
Compared to currently available urban flow datasets, the goal is to significantly increase the number of available samples while covering a broader range of building configurations. 
In this way, the 2D dataset presented in this work can serve as an efficient testbed for early-stage model development, while the upcoming 3D dataset will enable validation and application in more realistic urban environments.
Together, these datasets form a complementary foundation for the development of ML-based surrogate models for urban flow prediction, supporting both large-scale methodological research and future real-world applications.



\section*{Data availability}
The dataset described in this study is available on Havard Dataverse with the DOI: \href{https://doi.org/10.7910/DVN/J2DRQO}{10.7910/DVN/J2DRQO}.

\section*{Code availability}
The data-handling scripts can be accessed on Havard Dataverse with the DOI: \href{https://doi.org/10.7910/DVN/J2DRQO}{10.7910/DVN/J2DRQO}.

\section*{Acknowledgement}
This research was supported by the Korea Institute of Energy Technology Evaluation and Planning (KETEP) and the Ministry of Climate, Energy \& Environment (MCEE) of the Republic of Korea(RS-2023-00243974).
Furthermore, it was funded by the German Research Foundation within the Walter Benjamin fellowship RU~2771/1-1.
The authors gratefully acknowledge computing time on the supercomputer JURECA~\cite{JURECA} at Forschungszentrum Jülich under the compute project \textit{urbanflow}.

\section*{Author contributions}
H.L. created and maintains the dataset repository, developed the data handling codes, and analyzed the simulations.
A.L. acquired and supervised the computing resources and edited the manuscript. 
S.L. acquired funding,  provided advise from the fluid mechanics and \ac{ML} perspectives and edited the manuscript. 
M.R. acquired funding, acquired the computing resources, conducted and analyzed the simulations, and directed the project. 
H.L. and M.R. wrote the manuscript with contributions by A.L. and S.L.

\section*{Conflict of Interest statement}
The authors declare no competing interests.

\section*{References}

\bibliographystyle{abbrvurl}     
\bibliography{references}

\end{document}